\documentclass[range]{ar2e}

\usepackage{hyperref}
\usepackage{xcolor}
\usepackage{xspace}
\usepackage{graphicx}
\usepackage{subfigure}
\usepackage{rotating}
\usepackage{multirow}
\usepackage{mciteplus}
\usepackage{cite}

\usepackage{amsmath} 
\usepackage{amssymb}

\usepackage{ifthen} 
\newboolean{articletitles}
\setboolean{articletitles}{false} 
\newboolean{uprightparticles}
\setboolean{uprightparticles}{false} 

\def\babar  {\mbox{BaBar}\xspace}

\ifthenelse{\boolean{uprightparticles}}%
{

 \def\Pmu         {\ensuremath{\upmu}\xspace}

 \def\Ppi         {\ensuremath{\uppi}\xspace}                 
                  
 \def\Prho        {\ensuremath{\uprho}\xspace}                 
                  
 \def\Ptau        {\ensuremath{\uptau}\xspace}

 \def\Ppsi        {\ensuremath{\uppsi}\xspace}

 \def\PDelta      {\ensuremath{\Delta}\xspace}                 
 \def\PXi      {\ensuremath{\Xi}\xspace}                 
 \def\PLambda      {\ensuremath{\Lambda}\xspace}                 
 \def\PSigma      {\ensuremath{\Sigma}\xspace}                 
 \def\POmega      {\ensuremath{\Omega}\xspace}                 
 \def\PUpsilon      {\ensuremath{\Upsilon}\xspace}                 
 
 \def\PB      {\ensuremath{\mathrm{B}}\xspace}                 
                  
 \def\PD      {\ensuremath{\mathrm{D}}\xspace}

 \def\PJ      {\ensuremath{\mathrm{J}}\xspace}                 
 \def\PK      {\ensuremath{\mathrm{K}}\xspace}

 \def\PW      {\ensuremath{\mathrm{W}}\xspace}

 \def\PZ      {\ensuremath{\mathrm{Z}}\xspace}                 
                  
 \def\Pb      {\ensuremath{\mathrm{b}}\xspace}                 
 \def\Pc      {\ensuremath{\mathrm{c}}\xspace}                 
 \def\Pd      {\ensuremath{\mathrm{d}}\xspace}                 
 \def\Pe      {\ensuremath{\mathrm{e}}\xspace}

 \def\Pi      {\ensuremath{\mathrm{i}}\xspace}

 \def\Ps      {\ensuremath{\mathrm{s}}\xspace}                 
 \def\Pt      {\ensuremath{\mathrm{t}}\xspace}                 
 \def\Pu      {\ensuremath{\mathrm{u}}\xspace}

}
{

 \def\Pmu         {\ensuremath{\mu}\xspace}

 \def\Ppi         {\ensuremath{\pi}\xspace}                 
                  
 \def\Prho        {\ensuremath{\rho}\xspace}                 
                  
 \def\Ptau        {\ensuremath{\tau}\xspace}

 \def\Ppsi        {\ensuremath{\psi}\xspace}                 
                  
 \mathchardef\PDelta="7101
 \mathchardef\PXi="7104
 \mathchardef\PLambda="7103
 \mathchardef\PSigma="7106
 \mathchardef\POmega="710A
 \mathchardef\PUpsilon="7107
                  
 \def\PB      {\ensuremath{B}\xspace}                 
                  
 \def\PD      {\ensuremath{D}\xspace}

 \def\PJ      {\ensuremath{J}\xspace}                 
 \def\PK      {\ensuremath{K}\xspace}

 \def\PW      {\ensuremath{W}\xspace}

 \def\PZ      {\ensuremath{Z}\xspace}                 
                  
 \def\Pb      {\ensuremath{b}\xspace}                 
 \def\Pc      {\ensuremath{c}\xspace}                 
 \def\Pd      {\ensuremath{d}\xspace}                 
 \def\Pe      {\ensuremath{e}\xspace}

 \def\Pi      {\ensuremath{i}\xspace}

 \def\Ps      {\ensuremath{s}\xspace}                 
 \def\Pt      {\ensuremath{t}\xspace}                 
 \def\Pu      {\ensuremath{u}\xspace}

}

\def\en         {{\ensuremath{\Pe^-}}\xspace}   
\def\ep         {{\ensuremath{\Pe^+}}\xspace}
 
\def\epem       {{\ensuremath{\Pe^+\Pe^-}}\xspace}

\def\mup        {{\ensuremath{\Pmu^+}}\xspace}
\def\mun        {{\ensuremath{\Pmu^-}}\xspace} 
\def\mumu       {{\ensuremath{\Pmu^+\Pmu^-}}\xspace}

\def\tautau     {{\ensuremath{\Ptau^+\Ptau^-}}\xspace}

\def\Wpm    {{\ensuremath{\PW^\pm}}\xspace}
\def\Z      {{\ensuremath{\PZ}}\xspace}

\def\uquark    {{\ensuremath{\Pu}}\xspace}

\def\dquark    {{\ensuremath{\Pd}}\xspace}

\def\squark    {{\ensuremath{\Ps}}\xspace}

\def\cquark    {{\ensuremath{\Pc}}\xspace}
\def\cquarkbar {{\ensuremath{\overline \cquark}}\xspace}
\def\ccbar     {{\ensuremath{\cquark\cquarkbar}}\xspace}
\def\bquark    {{\ensuremath{\Pb}}\xspace}

\def\tquark    {{\ensuremath{\Pt}}\xspace}

\def\pion   {{\ensuremath{\Ppi}}\xspace}
\def\piz    {{\ensuremath{\pion^0}}\xspace}

\def\pip    {{\ensuremath{\pion^+}}\xspace}
\def\pim    {{\ensuremath{\pion^-}}\xspace}

\def\rhomeson {{\ensuremath{\Prho}}\xspace}
\def\rhoz     {{\ensuremath{\rhomeson^0}}\xspace}

\def\kaon    {{\ensuremath{\PK}}\xspace}
  \def\Kbar    {{\kern 0.2em\overline{\kern -0.2em \PK}{}}\xspace}

\def\Kp      {{\ensuremath{\kaon^+}}\xspace}
\def\Km      {{\ensuremath{\kaon^-}}\xspace}

\def\KS      {{\ensuremath{\kaon^0_{\rm\scriptscriptstyle S}}}\xspace}

\def\Kstarz  {{\ensuremath{\kaon^{*0}}}\xspace}
\def\Kstarzb {{\ensuremath{\Kbar{}^{*0}}}\xspace}
\def\Kstar   {{\ensuremath{\kaon^*}}\xspace}

\def\Kstarm  {{\ensuremath{\kaon^{*-}}}\xspace}

  \def\Dbar    {{\kern 0.2em\overline{\kern -0.2em \PD}{}}\xspace}
\def\D       {{\ensuremath{\PD}}\xspace}

\def\Dm      {{\ensuremath{\D^-}}\xspace}

\def\Dstarm  {{\ensuremath{\D^{*-}}}\xspace}

\def\Dsm     {{\ensuremath{\D^-_\squark}}\xspace}

\def\B       {{\ensuremath{\PB}}\xspace}
\def\Bbar    {{\ensuremath{\kern 0.18em\overline{\kern -0.18em \PB}{}}}\xspace}
\def\Bb      {{\ensuremath{\Bbar{}}}\xspace}
\def\Bz      {{\ensuremath{\B^0}}\xspace}
\def\Bzb     {{\ensuremath{\Bbar{}^0}}\xspace}
\def\Bu      {{\ensuremath{\B^+}}\xspace}

\def\Bp      {{\ensuremath{\Bu}}\xspace}

\def\Bd      {{\ensuremath{\B^0}}\xspace}
\def\Bs      {{\ensuremath{\B^0_\squark}}\xspace}
\def\Bds     {{\ensuremath{\B^0_{(\squark)}}}\xspace}
\def\Bsb     {{\ensuremath{\Bbar{}^0_\squark}}\xspace}
\def\Bdb     {{\ensuremath{\Bbar{}^0}}\xspace}
\def\Bdsb    {{\ensuremath{\Bbar{}^0_{(\squark)}}}\xspace}

\def\jpsi     {{\ensuremath{{\PJ\mskip -3mu/\mskip -2mu\Ppsi\mskip 2mu}}}\xspace}
\def\psitwos  {{\ensuremath{\Ppsi{(2S)}}}\xspace}

  \def\Y#1S{\ensuremath{\PUpsilon{(#1S)}}\xspace}

\def\Lz          {{\ensuremath{\PLambda}}\xspace}
\def\Lbar        {{\ensuremath{\kern 0.1em\overline{\kern -0.1em\PLambda}}}\xspace}

\def\Lb      {{\ensuremath{\Lz^0_\bquark}}\xspace}

\def\BF         {{\ensuremath{\cal B}}\xspace}

\def\BR         {\BF}
\newcommand{\decay}[2]{\ensuremath{#1\!\to #2}\xspace}         

\def\to                 {\ensuremath{\rightarrow}\xspace}

\def\qsq       {{\ensuremath{q^2}}\xspace}

\def\CP                {{\ensuremath{C\!P}}\xspace}

\def\AT#1     {\ensuremath{A_{\mathrm{T}}^{#1}}\xspace}           

\def\C#1      {\ensuremath{\mathcal{C}_{#1}}\xspace}                       
\def\Cp#1     {\ensuremath{\mathcal{C}_{#1}^{'}}\xspace}                    
\def\Ceff#1   {\ensuremath{\mathcal{C}_{#1}^{\mathrm{(eff)}}}\xspace}        
\def\Cpeff#1  {\ensuremath{\mathcal{C}_{#1}^{'\mathrm{(eff)}}}\xspace}       
\def\Ope#1    {\ensuremath{\mathcal{O}_{#1}}\xspace}                       
\def\Opep#1   {\ensuremath{\mathcal{O}_{#1}^{'}}\xspace}                    

\newcommand{\tev}{\ensuremath{\mathrm{\,Te\kern -0.1em V}}\xspace}
\newcommand{\gev}{\ensuremath{\mathrm{\,Ge\kern -0.1em V}}\xspace}
\newcommand{\mev}{\ensuremath{\mathrm{\,Me\kern -0.1em V}}\xspace}
\newcommand{\kev}{\ensuremath{\mathrm{\,ke\kern -0.1em V}}\xspace}
\newcommand{\ev}{\ensuremath{\mathrm{\,e\kern -0.1em V}}\xspace}
\newcommand{\gevc}{\ensuremath{{\mathrm{\,Ge\kern -0.1em V\!/}c}}\xspace}
\newcommand{\mevc}{\ensuremath{{\mathrm{\,Me\kern -0.1em V\!/}c}}\xspace}
\newcommand{\gevcc}{\ensuremath{{\mathrm{\,Ge\kern -0.1em V\!/}c^2}}\xspace}
\newcommand{\gevgevcccc}{\ensuremath{{\mathrm{\,Ge\kern -0.1em V^2\!/}c^4}}\xspace}
\newcommand{\mevcc}{\ensuremath{{\mathrm{\,Me\kern -0.1em V\!/}c^2}}\xspace}

\def\invfb   {\ensuremath{\mbox{\,fb}^{-1}}\xspace}

\def\deriv {\ensuremath{\mathrm{d}}}

\def\gsim{{~\raise.15em\hbox{$>$}\kern-.85em
          \lower.35em\hbox{$\sim$}~}\xspace}
\def\lsim{{~\raise.15em\hbox{$<$}\kern-.85em
          \lower.35em\hbox{$\sim$}~}\xspace}

\def\tell1  {TELL1\xspace}
\def\ukl1   {UKL1\xspace}

\newcommand{\eg}{\mbox{\itshape e.g.}\xspace}
\newcommand{\ie}{\mbox{\itshape i.e.}\xspace}

\newcommand{\etc}{\mbox{\itshape etc.}\xspace}

\usepackage{hyperref}    

\begin{document}

\jname{Annual Review of Nuclear and Particle Science}
\jyear{2015}
\jvol{65}
\ARinfo{1056-8700/97/0610-00}


\title{\boldmath Rare $b$ hadron decays at the LHC}

\markboth{{Blake,} {Gershon \&} {Hiller}}{Rare $b$ hadron decays at the LHC}

\author{
  T.~Blake and T.~Gershon
  \affiliation{Department of Physics, University of Warwick, Coventry, CV4 7AL, United Kingdom}
  G.~Hiller
  \affiliation{Institut f\"{u}r Physik, Technische Universit\"{a}t Dortmund, D-44221 Dortmund, Germany}
}


\begin{abstract}
  With the completion of Run~I of the CERN Large Hadron Collider, particle physics has entered a new era.
  The production of unprecedented numbers of heavy-flavoured hadrons in high energy proton-proton collisions allows detailed studies of flavour-changing processes.
  The increasingly precise measurements allow the Standard Model to be tested to a new level of accuracy.
  Rare $b$ hadron decays provide some of the most promising approaches for such tests, since there are several observables which can be cleanly interpreted from a theoretical viewpoint.
  In this article, the status and prospects in this field are reviewed, with a focus on precision measurements and null tests. 
\clearpage
\end{abstract}

\maketitle

\newpage

\section{Introduction}
\label{sec:intro}

Among the most distinctive features of the Standard Model (SM) of particle physics is the organisation of ``flavours'' of quarks and leptons.
Flavour-changes can occur only through the charged current weak interaction, so transitions between fermions of the same charge can only occur through loop processes~\cite{Glashow:1970gm}.
The probabilities of different transitions are governed by the elements of the appropriate fermion mixing matrices.
In particular, the fact that the Cabibbo-Kobayashi-Maskawa (CKM) quark-mixing matrix~\cite{Cabibbo:1963yz,Kobayashi:1973fv} is found to be approximately diagonal suppresses generation-changing transitions. 

Consequently, processes involving flavour changes between two up-type (\uquark, \cquark, \tquark) or between two down-type (\dquark, \squark, \bquark) quarks, \ie\ involving a flavour-changing neutral current (FCNC), occur only at loop level and are predicted to be rare within the SM.
Decays of \bquark~hadrons into final states containing a photon or a dilepton pair ($\epem$, $\mumu$) are of particular interest, and are the main topic of this review.
The rates and various kinematic distributions as well as \CP asymmetries, and other properties, of such rare decays can be predicted in the SM with low theoretical uncertainty, while the measured quantities may be affected by physics beyond the Standard Model (BSM), also referred to as ``New Physics'' (NP). 
Comparisons of the predictions with the measurements therefore provide
sensitive tests for BSM contributions.

This reason for interest in \bquark~hadron decays has been known since before the discovery of the \bquark~quark itself, and rare decays have been investigated by a number of experiments.
The discovery of the $b \to s\gamma$ process by the CLEO experiment~\cite{Alam:1994aw} has been followed by increasingly precise determinations culminating in results from the BaBar~\cite{Lees:2012ym,Lees:2012ufa} and Belle~\cite{Limosani:2009qg} experiments.
The consistency of these measurements with the latest theoretical prediction~\cite{Misiak:2006zs} provides strong constraints on BSM models.
Among the many other important results from the \B factory experiments, the first evidence for the $B^+ \to \tau^+ \nu_{\tau}$ decay~\cite{Aubert:2009wt,Lees:2012ju,Hara:2010dk,Adachi:2012mm,Kronenbitter:2015kls} is particularly germane to this discussion.  
The overall picture is one of consistency with the SM, but at a level of precision that mandates further experimental investigation.

The Large Hadron Collider (LHC) at CERN~\cite{Evans:2008zzb} provides the opportunity to make the next leap in precision.
Its high energy proton-proton collisions give a large cross-section of ${\cal O}(100\,\mu{\rm b})$~\cite{LHCb-PAPER-2010-002} for production of \bquark~quarks.
Due to the high luminosity delivered by the LHC, the decay products of the \bquark~hadrons that emerge from fragmentation are recorded in sufficient quantities to allow studies of rare decays at the ATLAS~\cite{Aad:2008zzm}, CMS~\cite{Chatrchyan:2008aa} and LHCb~\cite{Alves:2008zz} experiments.
For ATLAS and CMS, which instrument the central region of pseudorapidity, the online selection (``trigger'') requirements select only \bquark~hadron decays that contain a dimuon pair.
The LHCb detector, however, covers the forward region where \bquark~production peaks, and is designed to enable a broader range of \bquark~hadron decays, including those containing a photon or a dielectron pair, to be triggered; to achieve this LHCb must, however, operate at a lower instantaneous luminosity than the other experiments.
In the LHC Run~I data-taking period, corresponding to the calendar years (2011) 2012, when collisions were at centre-of-mass energies of (7) $8 \tev$, ATLAS and CMS each recorded approximately (5) $20\invfb$, while LHCb collected around (1) $2\invfb$.
These data samples contain unprecedented yields of numerous interesting rare decays of \bquark~hadrons, as will be discussed.

The focus of this review is the impact of the results, in the field of rare decays of \bquark~hadrons, from Run~I of the LHC.
This includes discussion of relevant results from other experiments, and a forward look to Run~II and beyond.
In order to find small deviations from the SM predictions, it is essential to aim for high precision, and therefore observables that can be both cleanly predicted and well measured are of greatest interest.
Such observables include relative and absolute rates, properties of kinematic distributions, and \CP asymmetries of decays involving a dilepton pair or a photon in the final state.
Certain processes that provide null tests of the SM, for example lepton flavour or lepton number violating decays, are also relevant in this context.
This selection of observables does not by far include all interesting measurements in quark flavour physics, or even in $B$ physics.
The interested reader is referred to reviews covering \CP violation in hadronic \bquark~hadron decays~\cite{HFAG,NirGershon,PDG2014}, the $\Bs$ system~\cite{Borissov:2013yha}, $D$ physics~\cite{Artuso:2008vf}, rare kaon decays~\cite{Bryman:2011zz,Cirigliano:2011ny} and top quark properties~\cite{Chierici:2014eqa,Deliot:2014uua}.
An earlier review on rare \bquark~hadron decays can be found in Ref.~\cite{Hurth:2010tk}.

The remainder of the review is organised as follows.
In Sec.~\ref{sec:theory} the theoretical framework is set out, while in Sec.~\ref{sec:expt} the experimental results are summarised.
These two aspects are brought together in Sec.~\ref{sec:interpretation} to enable interpretation of the results in the context of the SM and BSM theories.
A brief summary concludes the review in Sec.~\ref{sec:summary}.

\section{Theoretical framework}
\label{sec:theory}

The main challenges to develop the theory of rare $b$-decays in the LHC era are to improve precision of the predictions, and to perform and refine interpretations of the data in order to map the borders of the SM and possibly detail BSM features. 
The focus is on exclusive decays of \bquark~hadrons, including \Bs mesons and \bquark~baryons.
As regards the predictions, theory greatly benefits from the determination of crucial input such as quark mixing and masses from earlier experiments, and from maturing heavy quark methods for precision calculations of decay amplitudes.
The latter are based on the separation in energy scale between the mass of the $b$-quark and the energy scale of QCD ($m_b \gg \Lambda_{\rm QCD}$).
It is also possible to construct observables that are intrinsically robust against theoretical uncertainties, and hence provide precise tests of the SM with clean interpretation.
In this Section the framework for these tests is outlined.
In Sec.~\ref{sec:MIA}, the effective low energy Hamiltonian, whose induced couplings (the so-called ``Wilson coefficients'') can be used to describe the phenomenology of a wide range of decay modes, is introduced.
The status and recent advances of methods to determine QCD effects, in particular in exclusive $b \to s \ell^{+}\ell^{-}$ decays, are briefly reviewed in Sec.~\ref{sec:QCD}. 
In Sec.~\ref{sec:symmetry}, the optimised observables that are investigated experimentally are introduced, together with a discussion of consistency checks based on symmetry relations.
Finally, in Sec.~\ref{sec:NP} several explicit BSM theories, are discussed as examples of how deviations from the SM may appear in experiments.

\subsection{Model-independent analysis of $b \to s$ transitions}
\label{sec:MIA}

The large masses of the \Wpm, \Z and top quark compared to that of the beauty quark
allow the construction of a low energy effective field theory for $|\Delta B|=|\Delta S|=1$ transitions, with Hamiltonian
\begin{align}
  \label{eq:Heff}
  {\cal{H}}_{\rm eff} = 
   - \frac{4\, G_F}{\sqrt{2}} V_{tb}^{} V_{ts}^* \,\frac{\alpha_e}{4 \pi}\,
     \sum_i C_i(\mu_{\rm s}) {\cal{O}}_i(\mu_{\rm s}) \,,
\end{align}
where $G_F$ is the Fermi constant, $V_{ij}$ are CKM matrix elements and $\alpha_e$ is the fine structure constant. The $C_i(\mu_{\rm s})$ are Wilson coefficients corresponding to local operators with different Lorentz structure, ${\mathcal{O}}_i(\mu_{\rm s})$. The operators and their Wilson coefficients are evaluated at the renormalisation scale $\mu_{\rm s}$.
Doubly Cabibbo-suppressed contributions to the Hamiltonian $\propto V_{ub}^{} V_{us}^*$ have been neglected. 
Details of the effective Hamiltonian of Eq.~(\ref{eq:Heff}) can be found, for example, in Refs.~\cite{Buchalla:1995vs,Chetyrkin:1996vx}.

The following local operators are important for rare radiative, leptonic and semileptonic \bquark~hadron decays 
\begin{equation}\begin{array}{rcl@{\hspace{5mm}}rcl}     
  {\cal{O}}_{7}^{} & = & \frac{m_b}{e} \bar{s} \sigma^{\mu\nu} P_{R} b  F_{\mu\nu}\,, &
  {\cal{O}}_{7}^\prime & = & \frac{m_b}{e} \bar{s} \sigma^{\mu\nu} P_{L} b  F_{\mu\nu}\,, \\
  {\cal{O}}_{8}^{} & = & g_s \frac{m_b}{e^2} \bar{s} \sigma^{\mu\nu} P_{R} T^a b  G^a_{\mu\nu} \,, &
  {\cal{O}}_{8}^\prime & = & g_s \frac{m_b}{e^2} \bar{s} \sigma^{\mu\nu} P_{L} T^a b  G^a_{\mu\nu} \,, \\ 
  {\cal{O}}_{9}^{} & = &\bar{s} \gamma_\mu P_{L} b \, \bar{\ell} \gamma^\mu \ell \,, &
  {\cal{O}}_{9}^\prime & = & \bar{s} \gamma_\mu P_{R} b \, \bar{\ell} \gamma^\mu \ell \,, \\
  {\cal{O}}_{10}^{} & = & \bar{s} \gamma_\mu P_{L} b \, \bar{\ell} \gamma^\mu \gamma_5 \ell \,, &
  {\cal{O}}_{10}^\prime & = & \bar{s} \gamma_\mu P_{R} b \, \bar{\ell} \gamma^\mu \gamma_5\ell \,.
\end{array}\label{eq:operators}\end{equation}
Here, $P_{L/R}=(1 \mp \gamma_5)/2$ denotes a left/right handed chiral projection, $T^a$ represents the generators of QCD, and $F_{\mu \nu}$ ($G^a_{\mu \nu}$) is the electromagnetic (chromomagnetic) field strength tensor. 
The chirality-flipped operators $\mathcal{O}^\prime_{i}$ correspond to right-handed couplings and are obtained from the  $\mathcal{O}^{}_{i}$ by replacing $P_L \leftrightarrow P_R$. 
The left-handedness of the charged current interaction means that the Wilson coefficients $C_i^\prime$ corresponding to these primed operators are suppressed by ${\cal O}(m_s/m_b)$ in the SM.

The Wilson coefficients $C_i^{(\prime)}$ can be determined from measurements of observables in various different \bquark~hadron decay channels.
Among the operators of Eq.~(\ref{eq:operators}), radiative \bquark~hadron decays receive contributions from ${\cal O}^{(\prime)}_7$ and purely leptonic decays from ${\cal O}^{(\prime)}_{10}$.
Semileptonic $b \to s \ell^+\ell^-$ decays receive contributions from all of ${\cal O}^{(\prime)}_7$, ${\cal O}^{(\prime)}_9$ and ${\cal O}^{(\prime)}_{10}$. 
The $b \to d\gamma$ and $b \to d \ell^+\ell^-$ transitions are treated in an analogous way, but are further suppressed by $V_{tb}^{}V_{td}^{*}$ as opposed to $V_{tb}^{}V_{ts}^{*}$ in Eq.~(\ref{eq:Heff}). 
Consequently, in $b \to d$ transitions, \CP violation effects are generically larger because $V^{}_{ub} V^*_{ud}$ is of comparable magnitude to $V^{}_{tb} V^*_{td}$, though the stronger GIM-suppression of the $V^{}_{ub} V^*_{ud}$ term limits the size of any \CP asymmetry.
In the SM, scalar and pseudoscalar operators 
\begin{equation}\begin{array}{rcl@{\hspace{5mm}}rcl} 
  {\cal{O}}_{\rm S}^{} & = & \bar{s} P_{R} b \, \bar{\ell} \ell \, , &
  {\cal{O}}_{\rm S}^\prime & = & \bar{s} P_{L} b \, \bar{\ell} \ell \, , \\
  {\cal{O}}_{\rm P}^{} & = & \bar{s} P_{R} b \, \bar{\ell} \gamma_5 \ell \, , &
  {\cal{O}}_{\rm P}^\prime & = & \bar{s} P_{L} b \, \bar{\ell} \gamma_5 \ell \, 
\end{array}\end{equation}
are highly suppressed due to the small masses of the leptons, and can be safely neglected even for decays involving $\tau$ leptons. 
Contributions from tensor operators, 
\begin{equation}\begin{array}{rcl@{\hspace{5mm}}rcl} 
    {\cal{O}}_{\rm T}^{} & = & \bar{s} \sigma_{\mu\nu} b \, \bar{\ell} \sigma^{\mu\nu} \ell \, , & 
    {\cal{O}}_{\rm T5} & = &
     \bar{s} \sigma_{\mu\nu} b \, \bar{\ell} \sigma^{\mu\nu}\gamma_5 \ell \, ,
\end{array}\end{equation}
are also negligibly small in the SM.

The Wilson coefficients at the weak scale are obtained from matching amplitudes of the full electroweak theory onto ${\cal{H}}_{\rm eff}$. 
Below the $W$ mass, the Wilson coefficients follow renormalisation group evolution assuming SM dynamics~\cite{Buchalla:1995vs}.
The values at $\mu_{\rm s} = m_{b}$ are~\cite{Bobeth:1999mk} 
\begin{align}
  C_7^{\rm SM}  =-0.3 \, , \quad
  C_9^{\rm SM} =+4.2 \, , \quad  
  C_{10}^{\rm SM}  =-4.2 \, .
\end{align}
Comparisons of the measured values with the predictions provide sensitive tests of the SM.
BSM theories can modify the Wilson coefficients of ${\cal{H}}_{\rm eff}$, Eq.~(\ref{eq:Heff}), including those of operators not present or suppressed in the SM, $C_i^{(\prime)} = C_i^{(\prime)\,{\rm SM}} + C_i^{(\prime)\,{\rm NP}}$.
The number of possible new operators, at dimension six, is large and includes scalar, pseudoscalar and tensor operators.
If the BSM physics does not couple universally to leptons then sets of operators need to be considered separately for the different lepton flavours. 
New operators can also, in principle, induce lepton flavour-violating processes that are forbidden by accidental symmetries of the SM.  
Operators can also be associated with new sources of \CP violation, making their Wilson coefficients complex-valued. 
The large number of possible operators is intractable for a fully model-independent analysis. 
However, certain experimental signatures that can only be explained by particular extensions to the SM operator basis allow for a simplified analysis.

Recently it has become customary to rewrite the semileptonic operators of Eq.~(\ref{eq:Heff}) in a basis with left- and right- projected leptons~\cite{Alonso:2014csa,Hiller:2014yaa}
\begin{equation}\begin{array}{rcl@{\hspace{5mm}}rcl} 
  {\cal{O}}_{\rm LL}^{} & \equiv & ({\cal{O}}_9^{}-{\cal{O}}_{10}^{})/2 \,, &
  {\cal{O}}_{\rm LR}^{} & \equiv & ({\cal{O}}_9^{}+{\cal{O}}_{10}^{})/2 \,, \\  
  {\cal{O}}_{\rm RL}^{} & \equiv & ({\cal{O}}_9^{\prime}-{\cal{O}}_{10}^{\prime})/2 \,, & 
  {\cal{O}}_{\rm RR}^{} & \equiv & ({\cal{O}}_9^{\prime}+{\cal{O}}_{10}^{\prime})/2 \,,
\end{array}\end{equation}
where
\begin{equation}\begin{array}{rcl@{\hspace{5mm}}rcl} 
    C_{\rm LL}^{} & = & C_9^{}-C_{10}^{} \,, & 
    C_{\rm LR}^{} & = & C_9^{}+C_{10}^{} \,, \\
    C_{\rm RL}^{} & = & C_9^{\prime}-C_{10}^{\prime} \,, & 
    C_{\rm RR}^{} & = & C_9^{\prime}+C_{10}^{\prime} \,.
\end{array}\end{equation}
This basis change is useful in frameworks where BSM physics at a high mass scale respects the ${\rm SU}(2)_{\rm L}$ part of the SM gauge symmetry group, resulting in a simpler structure.
For instance, instead of fitting the two parameters $C^{}_9$ and $C^{}_{10}$, , if it is assumed that BSM physics contributes to ${\cal{O}}_{\rm LL}$ only, the constraint $C^{}_9+C^{}_{10}=0$ can be used.
In addition, ${\rm SU}(2)_{\rm L}$-relations between \bquark-decay observables and top physics can be obtained~\cite{Fox:2007in}.

\subsection{Non-hadronic $b$ hadron decays in QCD}
\label{sec:QCD}

\begin{figure}[!tb]
\centering
\includegraphics[width=0.85\linewidth]{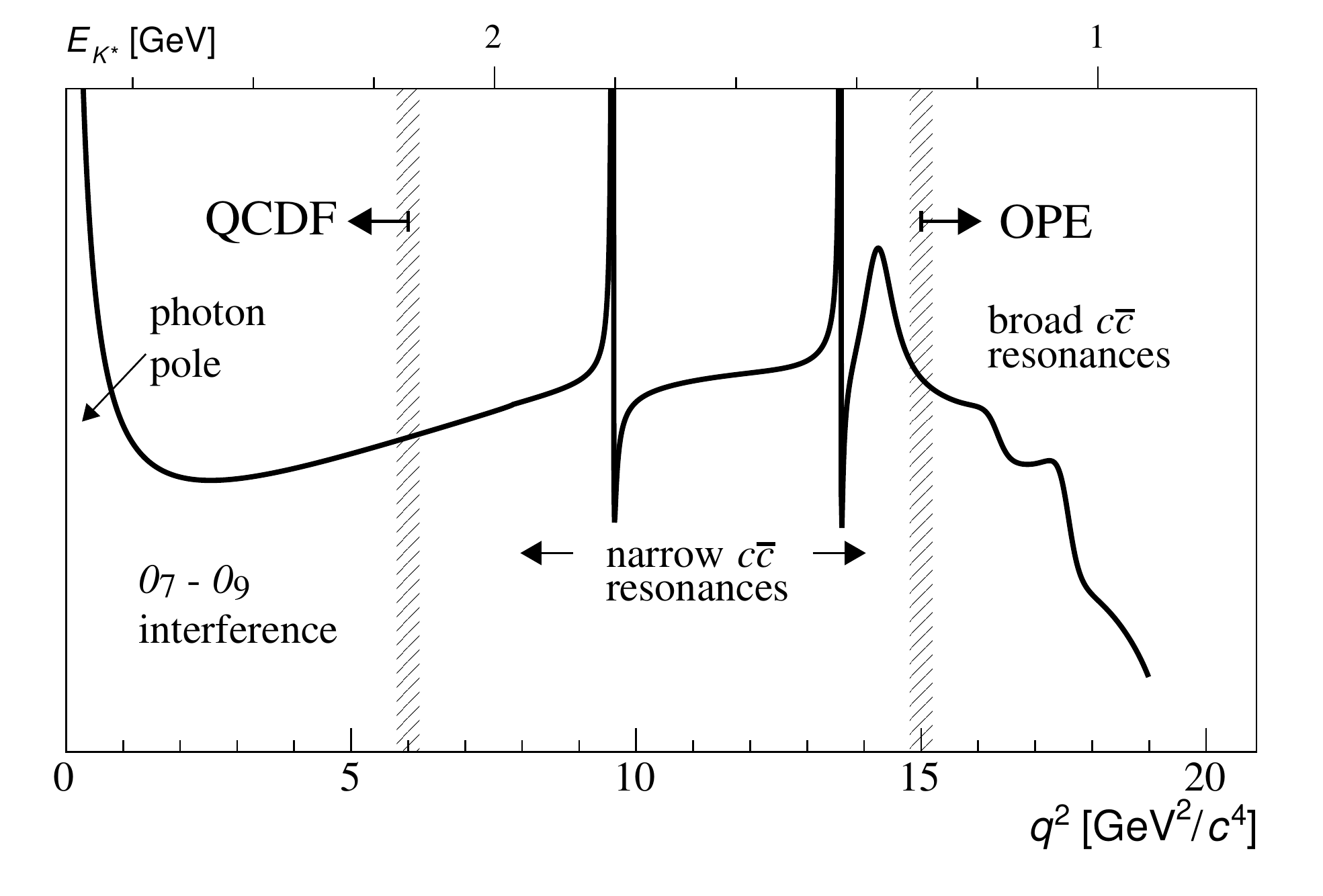}
\caption{
  Cartoon of the differential decay rate of \decay{\Bz}{\Kstarz\mumu} as a function of \qsq. 
  At very low \qsq (near maximal $E_{\Kstar}$), the virtual photon contribution associated to $C_7^{(\prime)}$ dominates. 
  As \qsq increases there is a region from $1 < \qsq < 6\gev^{2}/c^{4}$ where interference between $\mathcal{O}_{7}$ and $\mathcal{O}_{9}$ becomes large giving excellent sensitivity to NP in $C_9$. 
  At intermediate \qsq, the spectrum is dominated by the narrow \jpsi and \psitwos resonances. 
  At large \qsq (small $E_{\Kstar}$) contributions from broad charmonium resonances, above the open charm threshold, can be treated with a local OPE.
}
\label{fig:cartoon}
\end{figure}

Semileptonic heavy-to-light \bquark~hadron decays such as \decay{\Bd}{\Kstarz\mumu} have particularly interesting phenomenology.
These decays have sensitivity to electroweak physics in two kinematic regimes: at low invariant dilepton mass-squared ($\qsq$), where  the emitted hadron is energetic ($E \gg \Lambda_{\rm QCD}$ in the \bquark~hadron rest frame), QCD factorisation (QCDF) applies~\cite{Beneke:2001at}; at high invariant dilepton mass, the region of low hadronic recoil, where $\qsq ={\cal{O}}(m_b^2)$, an operator product expansion (OPE) in $1/m_b$ applies~\cite{Grinstein:2004vb}. 
These different kinematic regimes are indicated in Fig.~\ref{fig:cartoon}.
In both regimes, the heavy-to-light decays can be predicted systematically from QCD.
The methods to do this are now commonly employed, and in view of the experimental situation control of uncertainties becomes central. 
The dominant systematic uncertainties are parametric uncertainties from the hadronic transition form factors, $1/m_b$ power corrections (at low $\qsq$), and backgrounds from \ccbar resonances above the open charm threshold (at high $\qsq$).
In view of these issues it is mandatory to study the low and high $\qsq$ regions separately, and it has become conventional to perform analyses in finer bins of $\qsq$.

The transition form factors for heavy-to-light decays can be computed using the method of light cone sum rules (LCSR) if the final state hadronic system is energetic. 
Determinations for $B \to K$ and $B \to K^*$  form factors can be found in Refs.~\cite{Khodjamirian:2010vf,Khodjamirian:2012rm} and~\cite{Ball:2004rg}, respectively.  
Recently it has also been possible to determine the same form factors from lattice gauge theory. 
Lattice calculations are only applicable when the hadron is almost at rest in the \bquark~hadron decay frame, \ie\ at low recoil. 
The LCSR and lattice results are therefore complementary to each other, covering different kinematic regimes.
Unquenched lattice determinations can be found for $B \to K$~\cite{Bouchard:2013eph}, $B \to K^*$ and $\Bs \to \phi$~\cite{Horgan:2013hoa}, and $\Lb \to \Lz$~\cite{Detmold:2012vy} form factors. 
All lattice and most LCSR determinations assume that the light-particle being produced is both narrow and stable. 
This may not be a good approximation in some cases, in particular for the $B \to K^*$ transition due to the large width of the \Kstar\ resonance.
There are prospects for an improved treatment in future lattice calculations~\cite{Briceno:2014uqa}.

Issues of $1/m_b$ corrections exist for small $\qsq$ only because the power corrections at large \qsq are parametrically suppressed, bringing them to the few percent level.
The topic of power corrections has received a great deal of recent attention~\cite{Jager:2012uw,Descotes-Genon:2014uoa,Jager:2014rwa}. 
Eventually it will be possible to determine the corrections from data~\cite{Beaujean:2013soa}, or to subject them to consistency checks, as discussed in Sec.~\ref{sec:symmetry}.

\begin{figure}[!tb]
\centering
\includegraphics[width=0.7\linewidth]{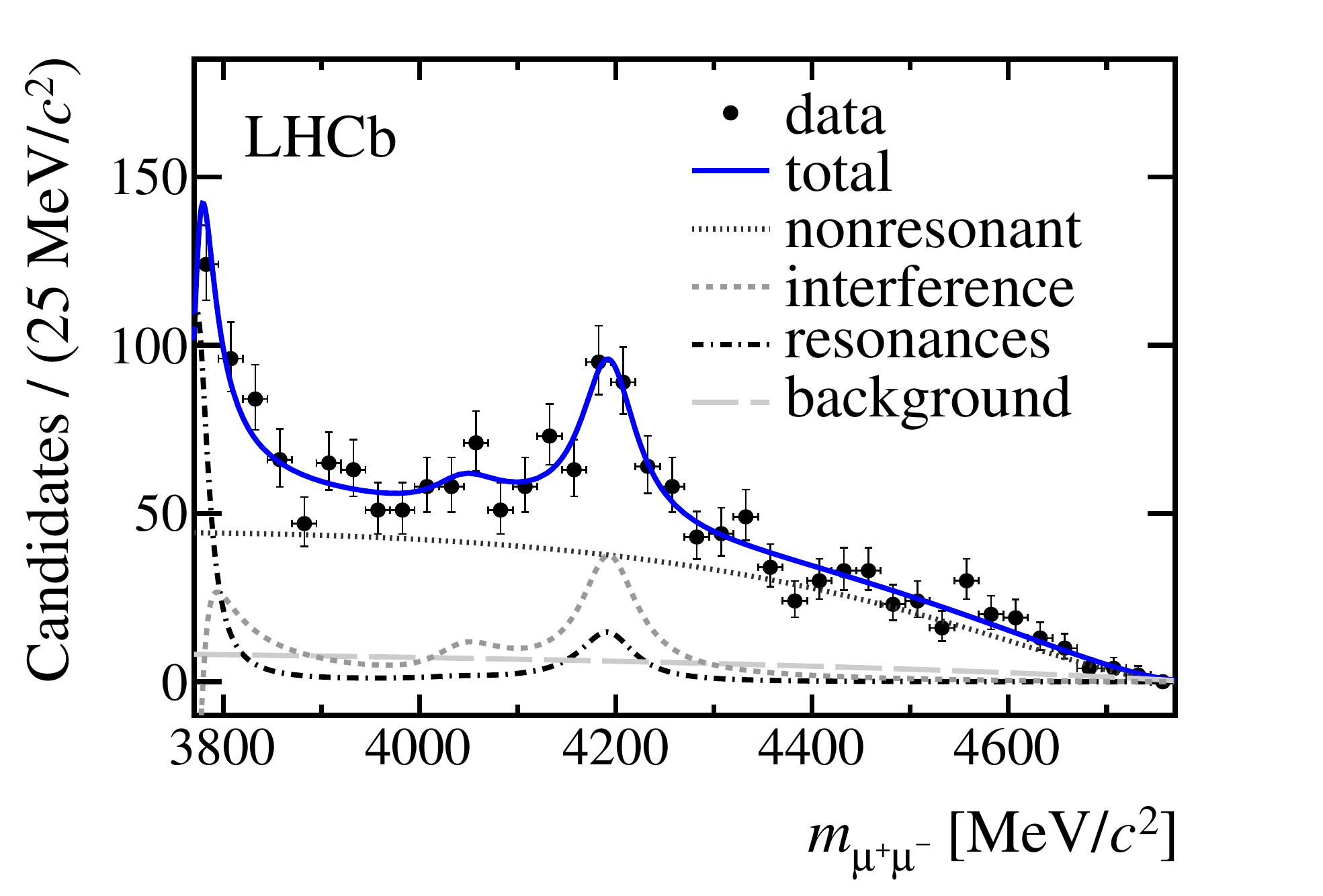}
\caption{
  Background subtracted dimuon mass distribution of \decay{\Bp}{\Kp\mumu} candidates~\cite{LHCb-PAPER-2013-039}. 
  Broad resonant contributions from the decays \decay{\Bp}{\psi(3770)\Kp} and \decay{\Bp}{\psi(4160)\Kp} are clearly visible.
}
\label{fig:exp:resonance}
\end{figure}

Amplitudes for rare semileptonic decays also receive contributions from the more prevalent \bquark~hadron decays to final states containing charmonia, through the quark-level transition $b \to c\bar{c}s$, where the $c\bar{c}$ resonance subsequently decays into dileptons.
In the effective theory, such contributions are induced by four-quark operators
\begin{equation}\begin{array}{rcl@{\hspace{5mm}}rcl} 
  {\cal{O}}_{1}^{} & = & \frac{4\pi}{\alpha_e} \bar{s} \gamma_\mu P_{L} b \, \bar c \gamma^\mu P_L c \, , &
  {\cal{O}}_{2}^{} & = & \frac{4\pi}{\alpha_e} \bar{s} \gamma_\mu P_{L} c \, \bar c \gamma^\mu P_L b \, .
\end{array}\end{equation}
These operators arise predominantly from tree-level $W$ exchange and have large, order one Wilson coefficients at the $b$-quark scale: $C_1\sim -0.2, \ C_2 \sim 1.1$. 
As shown in Fig.~\ref{fig:cartoon}, the most prominent effect of charmonia in semileptonic decays is the narrow resonance peaks at $\qsq = m_{\jpsi}^2, m_{\psi(2S)}^2$; these $\qsq$ regions have to be removed in experimental analyses.
In principle, however, the presence of charmonia affects the entire $\qsq$ region.
In radiative decays and at large recoil all charmonia are off-shell and are suppressed~\cite{Buchalla:1997ky,Khodjamirian:2010vf}. 
At high $\qsq$ a number of broad $c\bar{c}$ resonances can contribute~\cite{Kruger:1996cv,Ali:1999mm}. 
Such structure has been measured quite precisely in the \decay{\Bp}{\Kp\mumu} decay~\cite{LHCb-PAPER-2013-039}, as shown in Fig.~\ref{fig:exp:resonance}. 
Under a na\"{i}ve factorisation assumption, this structure can be compared to that from $e^+ e^- \to {\rm hadrons}$~\cite{Bai:2001ct,Ablikim:2007gd} using dispersion relations~\cite{Kruger:1996cv}.
Such a comparison~\cite{Lyon:2014hpa} revealed a dramatic deviation from expectation.
Although na\"{i}ve factorisation is not expected to be valid at high precision, this surprising difference needs to be understood in order to maximise the sensitivity to NP.

For completeness, it should be noted that there are also contributions from four-quark operators, referred to as ``QCD-penguins'', with flavour structure ${\cal{O}}_{3..6} \sim \bar s \gamma_\mu P_L b \sum_{u,d,s,c,b} \bar q \gamma^\mu P_{L,R} \, q$. 
Their SM Wilson coefficients at the $m_b$-scale are small, ${\cal O}(10^{-2})$, and therefore their effects in radiative and semileptonic \bquark~hadron decays are in general small.
Contributions from light resonances such as the $\phi$ meson, mediated by these operators, will however become important as the precision improves.

The OPE does not describe the resonance contributions locally in the amplitudes~\cite{Beylich:2011aq}.
It is, however, expected to capture their effect after integrating over a sufficient range of $\qsq$.
It is therefore important to investigate the optimal binning for precision BSM searches. 
As the OPE does predict universality between certain transversity amplitudes, cancellation of the effects of resonances is expected in particular ratios~\cite{Hiller:2013cza}.

\subsection{Optimised observables and symmetry relations}
\label{sec:symmetry}

Semileptonic $b \to s\ell^+\ell^-$ decays with a vector meson in the final state provide a particularly rich set of observables that can be accessed through the angular distribution of the particles in the decay. 
The decay rate for these semileptonic processes can be expressed in terms of decay amplitudes for the vector meson corresponding to different transversity states, $A^{\rm L,R}_{\parallel}$, $A^{\rm L,R}_{\perp}$ and $A^{\rm L,R}_{0}$. 
The axial-vector coupling through $C_{10}$ allows to distinguish between amplitudes with a left- and right-handed chirality of the dilepton system.
For completeness, it should be noted that a fourth amplitude ($A_t$) exists corresponding to the spin-0 $\ell^+\ell^-$ configuration.
The effect of this amplitude is however suppressed by the small lepton mass.
In BSM models with (pseudo)-scalar operators the effect of $A_t$ can be enhanced, and a further amplitude, $A_S$, appears.
In the presence of tensor operators, six additional transversity amplitudes can enter~\cite{Bobeth:2012vn}.

It is then possible to construct angular observables that are related to these decay amplitudes, for example the fraction of longitudinal polarisation of the vector meson, 
\begin{equation}
F_{\rm L} = \frac{|A_{0}^{\rm L}|^{2} +  |A_{0}^{\rm R}|^{2}}{|A_{0}^{\rm L}|^{2} +  |A_{0}^{\rm R}|^{2} + |A_{\parallel}^{\rm L}|^{2} +  |A_{\parallel}^{\rm R}|^{2} + |A_{\perp}^{\rm L}|^{2} +  |A_{\perp}^{\rm R}|^{2}}~. 
\end{equation}
One of the most widely discussed observables is the forward-backward asymmetry of the dilepton system, 
\begin{equation}
A_{\rm FB} = \frac{4}{3} \frac{{\rm Re} \left(A_{\parallel}^{\rm L} A_{\perp}^{\rm L*} - A_{\parallel}^{\rm R} A_{\perp}^{\rm R*}\right)}{|A_{0}^{\rm L}|^{2} +  |A_{0}^{\rm R}|^{2} + |A_{\parallel}^{\rm L}|^{2} +  |A_{\parallel}^{\rm R}|^{2} + |A_{\perp}^{\rm L}|^{2} +  |A_{\perp}^{\rm R}|^{2}}~,
\end{equation} 
which arises from the chiral nature of the coupling to the leptons and flips sign at $\qsq \approx 4\gev^{2}/c^{4}$~\cite{Beneke:2001at,Bobeth:2011nj,Kumar:2014bna} due to interference between the photon dipole ($\mathcal{O}^{(\prime)}_{7}$) and vector operators ($\mathcal{O}^{(\prime)}_{9}$). 

It is also possible to build sets of ``optimised'' observables from the decay amplitudes in which the leading form factor uncertainties cancel. 
The design of such observables is based on the $1/m_b$ expansion. 
In both kinematic regions of interest (low \qsq and high \qsq) the transversity amplitudes receive their leading contribution from a factorisable form factor term
\begin{equation}
 A^{\rm L,R}_i = f_i \times C_i^{\rm L,R} + \text{non fact.}\, , \quad i=0,\parallel, \perp \, ,
\end{equation}
where the $f_i$ are corresponding transversity form factors and contain long-distance (QCD) information only. 
The $C_i^{\rm L,R}$ on the other hand denote short-distance coefficients, sensitive to electroweak scale physics. 
They are composed of Wilson coefficients of the semileptonic four-fermion operators as well as contributions from four-quark and dipole operators (see Sec.~\ref{sec:MIA}).

At low $\qsq$, the form factor relations imply that $f_\parallel = f_\perp + \mathcal{O}(1/m_b)$, such that $A_{\parallel}^{\rm L,R} \approx -A_{\perp}^{\rm L,R}$ if $C^\prime = 0$. 
This relationship allows to construct SM null tests with sensitivity to right handed currents, for example~\cite{Kruger:2005ep}
\begin{equation}
  \label{eq:AT2}
  A_{\rm T}^{(2)}  = \frac{|A^{\rm L}_{\perp}|^{2} + |A^{\rm R}_{\perp}|^{2} - |A^{\rm L}_{\parallel}|^{2} - |A^{\rm R}_{\parallel}|^{2}}{|A^{\rm L}_{\perp}|^{2} + |A^{\rm R}_{\perp}|^{2} + |A^{\rm L}_{\parallel}|^{2}  + |A^{\rm R}_{\parallel}|^{2}}~,
\end{equation}
which is expected to be very close to zero in the SM.
It is also possible to build other clean observables from bilinear combinations of amplitudes such that $f_{\parallel}$, $f_{\perp}$ and $f_{0}$ cancel at leading order in $1/m_b$. 
The $P^\prime$-family~\cite{DescotesGenon:2012zf} is a good example of a set of these clean observables, \eg
\begin{equation}
\begin{split}
P^\prime_5 &= \sqrt{2} \frac{{\rm Re}\left(A_{0}^{\rm L} A_{\perp}^{\rm L*} - A_{0}^{\rm R} A_{\perp}^{\rm R*}\right)}{\sqrt{\left(|A_{0}^{\rm L}|^{2} +  |A_{0}^{\rm R}|^{2}\right) \left(|A_{\parallel}^{\rm L}|^{2} +  |A_{\parallel}^{\rm R}|^{2} + |A_{\perp}^{\rm L}|^{2} +  |A_{\perp}^{\rm R}|^{2}\right)}} \, , \\
& \approx \sqrt{2} \frac{{\rm Re}\left( C_0^{\rm L} C_\perp^{\rm L} - C_0^{\rm R} C_\perp^{\rm R} \right) }{\sqrt{ \left( |C_0^{\rm L}|^{2} + |C_0^{\rm L}|^{2} \right) \left( |C_{\perp}^{\rm L}|^{2} + |C_{\perp}^{\rm L}|^{2}  + |C_{\parallel}^{\rm L}|^{2} + |C_{\parallel}^{\rm L}|^{2} \right) }} \, .
\end{split} 
\end{equation} 

At large \qsq, the OPE predicts a different kind of relationship 
\begin{equation}
\begin{split}
 A^{\rm L,R}_{\parallel,0} & =f_{\parallel,0} \times C_-^{\rm L,R}  \, , \quad  A^{\rm L,R}_\perp=f_\perp C_+^{\rm L,R} \, , \\
 C_-^{\rm L,R} &= C_+^{\rm L,R} ~~{\textrm{if and only if}} ~~C^\prime=0 \, .
\end{split} 
\end{equation} 
It follows immediately that the ratio $A^{\rm L,R}_0/A^{\rm L,R}_\parallel$ is short-distance-free, \ie\ is independent of $C_i$. 
This feature allows to extract the form factor ratio $f_0/f_\parallel$ directly from data.
If there are no right-handed currents, $C^\prime = 0$, the form factor ratio $f_\perp/f_\parallel$ can be extracted in this kinematic region from $A_{\rm T}^{(2)}$. 
This information about the hadronic system, can in turn be used to provide better control of uncertainties on other observables.
It is also possible to derive form factor free observables at large \qsq, such as the $H_{\rm T}^{(i)}$ family~\cite{Bobeth:2010wg}. 

At zero recoil, when $\qsq=(m_B-m_{K^*})^2$, there are exact relationships between the amplitudes due to ambiguity of the direction of the \Kstar in the \B rest frame. 
At this kinematic end-point, $A^{\rm L, R}_{\perp} = 0$ and $A^{\rm L, R}_{\parallel} = -\sqrt{2}A^{\rm L, R}_{0}$~\cite{Hiller:2013cza}. 
The endpoint predictions can be compared to data, as shown in Table~\ref{tab:vergleich-all}, to provide a consistency check.
Parity selection allows to extend the prediction to the vicinity of the endpoint, where observables such as $A_{\rm FB}$ and $P^\prime_5$ vary linearly with the modulus of the $\Kstar$ three momentum in the $B$ centre-of-mass frame, and the slope provides a test of the SM.

\begin{table}[!tb]
\centering
\caption{
  Data on $B \to K^* \ell^+ \ell^-$  and $\Bs \to \phi \ell^+ \ell^-$ in the endpoint-bin $\qsq \in [16,19]\, \mbox{GeV}^2$ (LHC experiments) or otherwise $\qsq \in [16 \, \mbox{GeV}^2, \qsq_{\rm max}]$ versus the endpoint prediction.
  Note that $S_3=1/2 (1-F_L) A_{\rm T}^{(2)}$.  
  Adapted from Ref.~\cite{Hiller:2013cza}.
}
\label{tab:vergleich-all}
\resizebox{\textwidth}{!}{
\begin{tabular}{ l | c c c c  c c  }
 & $F_L$  & $S_3$  & $P_4^\prime$ & $S_7$ & $ P_5^\prime/A_{\rm FB}$ & $S_8/S_9$   \\ \hline
Endpoint & $1/3$ & $-1/3$ & $ \sqrt{2}$ & $ 0 $  & $\sqrt{2}$ & $-1/2$  \\ \hline
$\B \to \Kstar \ell^+ \ell^-$ & $0.38 \pm 0.04$ & $-0.22 \pm 0.09$ & $0.70\,^{+0.44}_{-0.52}$ & $0.15\,^{+0.16}_{-0.15}$ & $1.63 \pm 0.57$ & $-0.5 \pm 2.2$  \\
$\Bs \to \phi \ell^+ \ell^-$ & $0.16\,^{+0.18}_{-0.12}$ & $0.19\,^{+0.30}_{-0.31} $ & -- & --  & -- & -- \\
\end{tabular}
}
\end{table}

There are also more general relationships between the angular observables $J_i$ of the $B \to K^* \ell^+\ell^-$ decay, defined in Sec.~\ref{sec:expt:kstarll}, due to the composition of the observables in terms of pairs of  transversity amplitudes. 
These relationships, which are valid over the whole range of $\qsq$, serve as a further consistency check of the experimental results~\cite{Matias:2014jua}.

\subsection{Benchmarking NP}
\label{sec:NP}

An important motivation when building BSM models is the origin of electroweak symmetry breaking and the stabilisation of the weak scale, \ie\ the so-called ``hierarchy problem'' (see \eg\ Ref.~\cite{Feng:2013pwa}).
Among the most promising approaches are models that invoke supersymmetry, extra dimensions, new strong interactions or combinations thereof. 
Generically, in these models flavour violation beyond the CKM matrix is induced. 
Severe constraints on masses and couplings must be imposed for the NP to maintain contact with the electroweak scale:
without any flavour suppression (which could be SM-like, with GIM- and CKM-like effects) then NP is pushed up to scales as high as $10^5\tev$~\cite{Isidori:2010kg}.
In turn, the non-observation of BSM effects in the flavour sector provides important information about NP.

\subsubsection*{Minimal flavour violation}

Standard Model extensions can be classified according to their amount of flavour violation. 
The term {\it minimal flavour violation} (MFV) is used for models where flavour is broken in an SM-like way.
Formally, within MFV the spurion fields which break the flavour symmetry that would be present in the SM in the absence of quark masses correspond to the SM Yukawa couplings~\cite{D'Ambrosio:2002ex}.
Hence the flavour violation can be parametrised in terms of quark masses and CKM elements, which are known parameters of the SM.
Any deviation from MFV corresponds to NP.
The MFV paradigm provides an attractive way to resolve the tension between the expectation that the NP scale should be ${\cal O}(1\tev)$ due to naturalness arguments, 
while limits from FCNC processes assuming generic NP couplings point to a much higher scale. 
Still, viable non-MFV models exist with BSM around the \tev scale or higher.

The MFV framework can be tested through \CP violating observables as well as rare decays.
Since, as discussed in Sec.~\ref{sec:MIA}, the effective Hamiltonians for $b \to d$ and $b \to s$ transitions share the same structure, ratios of $b \to d$ and $b \to s$ processes provide powerful tests of MFV.
Generically, such ratios can be predicted in the SM to be equal to $|V_{td}/V_{ts}|^2 \approx 1/25$ modified by hadronic matrix elements and phase space factors, while in non-MFV models they can take very different values.
One important example is the ratio of \Bd and \Bs dimuon decay rates~\cite{Bobeth:2002ch}, but $\BR(\decay{\Bz}{\rho\gamma})/\BR(\decay{\Bz}{\Kstarz\gamma})$~\cite{Ali:2004hn,Ball:2006eu} and $\BR(\decay{\Bp}{\pip \ell^+ \ell^-})/\BR(\decay{\Bp}{\Kp \ell^+ \ell^-})$~\cite{Ali:2013zfa} exhibit similar features.
To reach high precision, good control of ${\rm SU}(3)$-breaking in the hadronic matrix elements is required.
The ratio $\BR(\decay{\Bs}{\Kstarzb \ell^+ \ell^-})/\BR(\decay{\Bz}{\Kstarz \ell^+ \ell^-})$ may provide a complementary approach to control such uncertainties.

\subsubsection*{Model building and simplified models}

In addition to modifying FCNCs, NP models typically predict new particles that can be searched for at the LHC.
As evidence for these new particles has not been found in Run~I, limits on the masses of these particles have been pushed into the $\tev$ range~\cite{Halkiadakis:2014qda}.
To keep the relation to the weak scale, without invoking any fine tuning of parameters, new directions of model building have appeared, \eg\ Ref.~\cite{1126-6708-2005-06-073,Giudice200465,Fan:2011yu}.

Instead of building models that are complete up to the GUT or Planck scale, it is common to consider simplified models. 
These usually comprise the SM plus one new sector with a rather small number of new parameters, making them predictive and easy to constrain. 
The choice of new sector that is added is either made by theoretical prejudice or is driven by a need to explain a deficiency of the SM or a discrepancy between SM predictions and data.
Two such simplified models, motivated by current hints of discrepancies in the $B \to K^{(*)} \ell^+ \ell^-$ data, are $Z$-penguins and leptoquarks. 

A $Z$-penguin is a FCNC involving a neutral external field that originates from a ${\rm U}(1)$ gauge interaction. 
In the case of the SM, $Z$-penguins arise at loop-level and are induced by the weak interaction. 
Modifications to the effective couplings arise generically in most SM extensions~\cite{Buchalla:2000sk}.
For the $b \to s$ transition,
\begin{align}
\mathcal{L}^Z=Z^\mu (g_{sb}^L \bar s \gamma_\mu P_L b + g_{sb}^R \bar s \gamma_\mu P_R b) + {\rm h.c.} \, ,
\end{align}
where the couplings $g_{sb}^{L/R}$, which can be related to the Wilson coefficients $C_{9,10}^{(\prime)}$, are generically complex. 
If the couplings to the leptons are SM-like, the vector current coupling is suppressed relative to the axial-vector one by a factor $\left| 4 \sin^2 \theta_W - 1 \right|$, where $\theta_W$ is the weak mixing angle, and the main contribution is through $C_{10}^{(\prime)}$.
With ${\rm U}(1)$-extensions of the SM ($Z^\prime$ models), however, the contribution to $C_{9}^{(\prime)}$ can also be sizable.
An example of a $Z^\prime$ model is a gauge extension to $\tau-\mu$ lepton number~\cite{Fox:2011qd,Altmannshofer:2014cfa}. 
A survey of the parameter space of $Z^\prime$ models can be found in Ref.~\cite{Buras:2014zga}.

Leptoquarks are bosonic particles that carry one lepton and one quark flavour quantum number. 
They can be spin one but are more commonly assumed to be scalar particles ($\phi$) which have Yukawa-like couplings ($\lambda_{q \ell}^L$, $\lambda_{q \ell}^R$) to the (left- or right-handed, respectively) SM fermions,
\begin{align}
\label{eq:eyukawa}
\mathcal{L}^{\rm LQ}=-\lambda_{q \ell}^L\, \phi\, (\bar{q} P_L \ell)  -\lambda_{q \ell}^R\, \phi\, (\bar{q} P_R \ell) + {\rm h.c.}
\end{align}
Tree-level $\phi$-exchange induces processes such as $b \to (s,d) \ell\ell$, which, depending on the handedness of the interaction, results in a modification of the semileptonic Wilson coefficients $C_{9,10}^{(\prime)}$~\cite{Hiller:2014yaa}. 
Leptoquarks can also provide a natural explanation for non-universal couplings to leptons in $b \to s \ell^+ \ell^-$ processes. 
Generally, leptoquarks also induce lepton flavour violation, requiring an extension to the SM operator basis. 
Limits on $\BR(\mu^{\pm} \to e^{\pm} \gamma)$ and $\BR(\Bds \to e^\pm\mu^\mp)$ strongly constrain the couplings involving electrons and muons, but the parameter space for other couplings remains viable.
With benchmark masses between 1--50\tev, motivated by the hint of lepton nonuniversality discussed in Sec.~\ref{sec:expt:luv}, it would be challenging to observe directly produced leptoquarks at the LHC but effects could be visible in rare decay processes such as $B \to K \tau^\pm \mu^\mp$ and $\Bds \to \tau^\pm \mu^\mp$~\cite{Davidson:1993qk,Glashow:2014iga,Varzielas:2015iva}.

\section{Status and prospects of measurements}
\label{sec:expt}

\subsection{Dilepton decays}
\label{sec:expt:dimuon}

The leptonic \decay{\Bs}{\mumu} and \decay{\Bz}{\mumu} decays are exceedingly rare in the SM. 
In addition to being loop- and CKM- suppressed, the decay of a pseudoscalar \B meson into a pair of muons has significant helicity suppression in the SM. 
The SM values of the time-integrated branching fractions~\cite{DeBruyn:2012wj} can be expressed as~\cite{Buras:2012ru,Bobeth:2013uxa} 
\begin{equation}
  \overline{\cal B}(\decay{\Bds}{\mumu}) = \frac{\left| V_{tb}^* V_{tq}^{} \right|^2 G_F^2 \; \alpha^{2}_{e} \; M_{\Bds} M_\mu^2 f_{\Bds}^2}{16 \pi^3 \Gamma_{q{\rm H}}} \sqrt{1 - \frac{4M_\mu^2}{M_{\Bds}^2}} \left| C_{10}(m_b) \right|^2 + ...
\end{equation} 
where the ellipses denote subleading terms.
In the above, $M$ denotes the mass of the particle in subscript and $\Gamma_{q{\rm H}}$ is the total width of the heavier of the two mass eigenstates in the $\Bds$--$\Bdsb$ system.
While those quantities are all well-known from experiments, the decay constant $f_{\Bds}$ must be determined from lattice QCD.
Using values of $f_{\Bds}$ obtained by averaging different calculations~\cite{Aoki:2013ldr,Bazavov:2011aa,McNeile:2011ng,Na:2012kp} and including higher-order QCD and EW corrections~\cite{Hermann:2013kca,Bobeth:2013tba}, the latest SM predictions are~\cite{Bobeth:2013uxa} 
\begin{equation}
\begin{split}
\overline{\cal B}(\decay{\Bs}{\mumu}) & = (3.65\pm 0.23)\times 10^{-9} ~~{\rm and} \\
\overline{\cal B}(\decay{\Bz}{\mumu}) & = (1.06\pm 0.09)\times 10^{-10}  ~. \\ 
\end{split}
\end{equation} 
The uncertainties on the SM predictions mainly come from the knowledge of the decay constants and the CKM matrix elements. 
In both cases, improvement can be anticipated with refined lattice QCD calculations.

The suppression of the \decay{\Bds}{\mumu} branching fractions is characteristic of the SM.
In particular, scalar contributions from SM Higgs penguin diagrams are negligible due to the small size of the muon mass. 
However, many BSM models can cause the branching fractions to deviate from their SM values.
Models with extended Higgs' sectors, for example, can produce significant
enhancements in the rates of the decays as the helicity suppression is broken
(see, for example, Refs.~\cite{Huang:1998vb,Choudhury:1998ze,Babu:1999hn,Bobeth:2001sq}).
The ratio of branching fractions for \Bz and \Bs decays to dimuons also provides a stringent test of MFV.

Prior to data taking at the LHC, no evidence for either decay had been found and limits on their branching fractions were still an order of magnitude above the SM expectations~\cite{Abazov:2010fs,Aaltonen:2011fi}. 
A series of results from LHC experiments~\cite{LHCb-PAPER-2011-004,LHCb-PAPER-2011-025,LHCb-PAPER-2012-007,Chatrchyan:2012rga,Aad:2012pn} significantly restricted the available phase space for BSM theories, giving strong constraints complementary to those from searches for on-shell production of new particles.
In summer 2013 both CMS and LHCb were both able to report evidence for the \decay{\Bs}{\mumu} decay at the level of four standard deviations ($\sigma$) using their full Run~I datasets~\cite{Chatrchyan:2013bka,LHCb-PAPER-2013-046}.
The experiments exploit multivariate event classifiers to optimise the separation of signal from backgrounds consisting of muons from different $b$-hadron decays. 
Backgrounds from $b$-hadron decays where one or more particles were mistakenly identified as muon or where one or more particles from a $b$-hadron decay was not reconstructed were estimated using samples of simulated events whose performance had been corrected to match that of data. 

\begin{figure}[!tb]
\centering
\begin{minipage}[c]{0.66\textwidth}
\includegraphics[width=\linewidth]{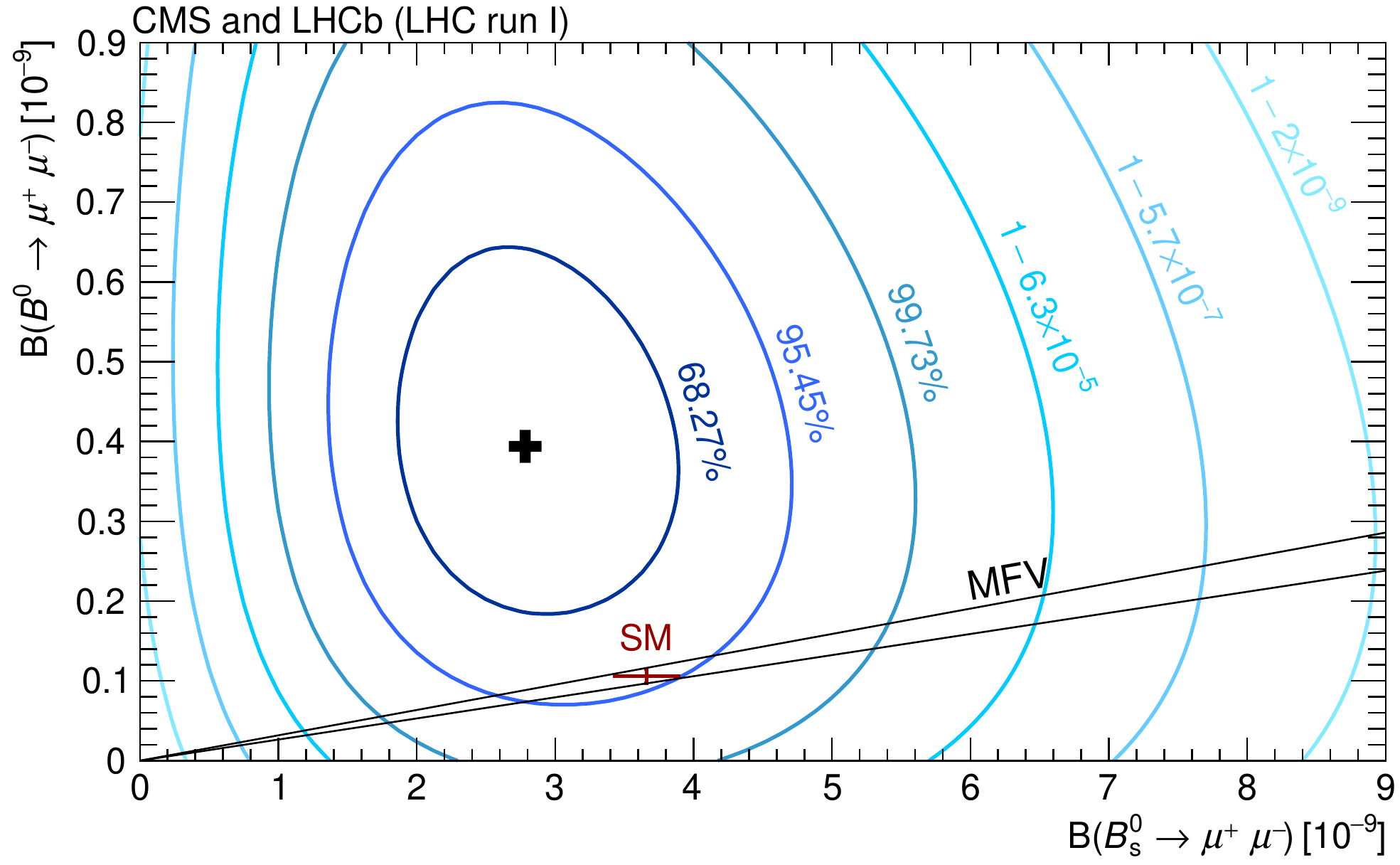} 
\end{minipage}
\begin{minipage}[c]{0.31\textwidth} 
\includegraphics[width=\linewidth]{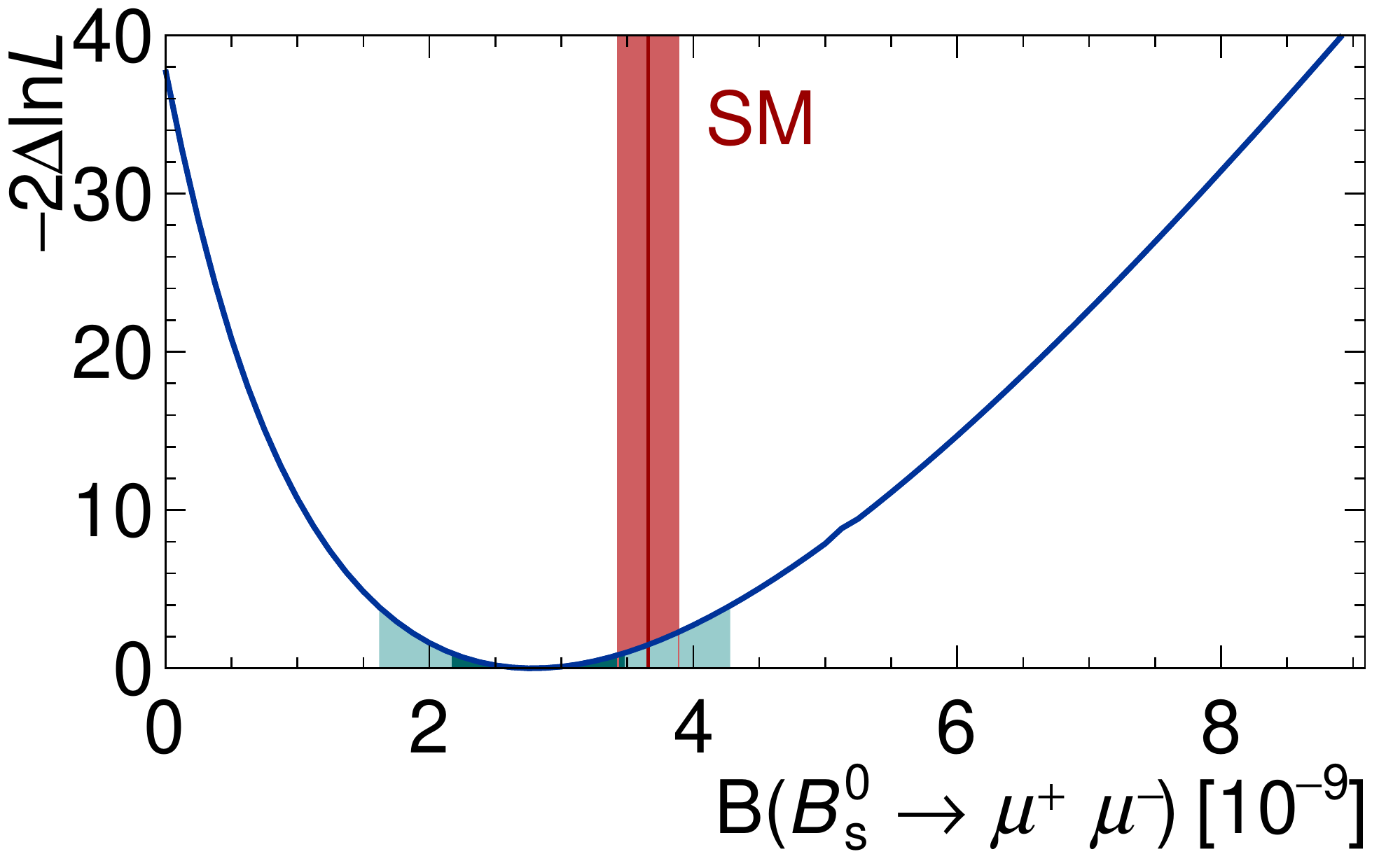} \\ 
\includegraphics[width=\linewidth]{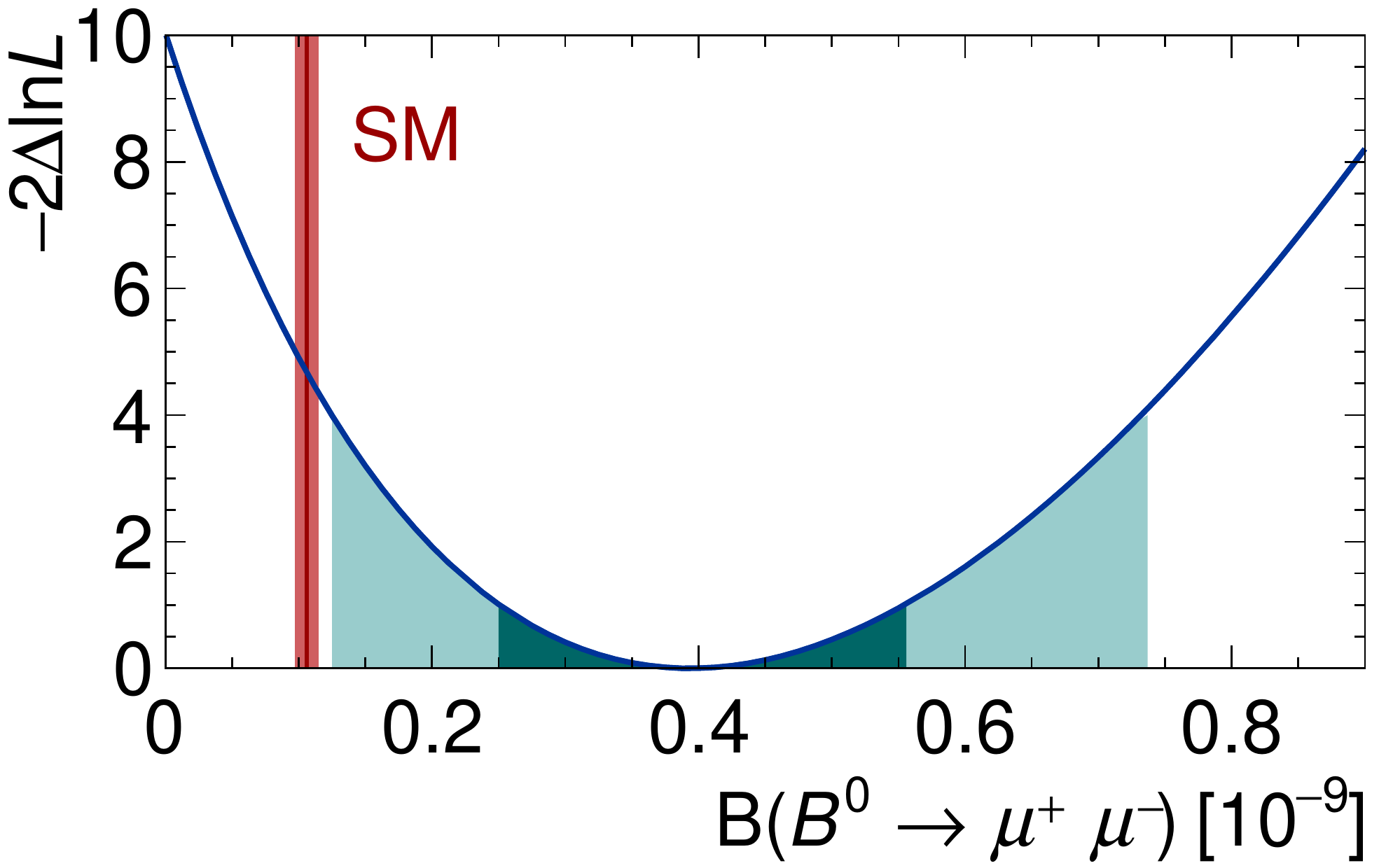}
\end{minipage}
\caption{
  Likelihood contours for the \decay{\Bs}{\mumu} and \decay{\Bz}{\mumu} branching fractions from a fit to the combined LHCb and CMS datasets.
  The cross indicates the best fit point in the two-dimensional plane of the branching fractions, where the SM and MFV predictions are also shown.
  The one dimensional projections of the likelihood are also shown.
  Modified from Ref.~\cite{LHCb-PAPER-2014-049}.
\label{fig:Bsmumu}
}
\end{figure}

To obtain the best information from the LHC Run~I dataset, a simultaneous analysis of the CMS and LHCb datasets was performed, giving~\cite{LHCb-PAPER-2014-049}
\begin{equation}
\begin{split}
\overline{\cal B}(\decay{\Bs}{\mumu}) & =   (2.8\,^{\,+0.7}_{\,-0.6}) \times 10^{-9} ~~{\rm and} \\
\overline{\cal B}(\decay{\Bz}{\mumu}) & =   (3.9\,^{\,+1.6}_{\,-1.4}) \times 10^{-10} ~. \\
\end{split}
\end{equation}
These measurements constitute the first observation of the \decay{\Bs}{\mumu} decay at more than $6\sigma$ and the first evidence for the \decay{\Bz}{\mumu} decay at more than $3\sigma$. 
The results, shown in Fig.~\ref{fig:Bsmumu}, are compatible with the SM at the level of around $2\sigma$ and put strong constraints on possible BSM scalar and pseudoscalar operators as discussed in Sec.~\ref{sec:interpretation}.

The observation of the \decay{\Bs}{\mumu} decay and the first evidence for the \decay{\Bz}{\mumu} decay represent the culmination of an experimental search that lasted three decades.
The progress for the other dilepton modes has been somewhat less dramatic -- as summarised in Table~\ref{tab:b2ll}, the branching fraction limits are at least five orders of magnitude above the SM expectations.  
Prospects for improved measurements are discussed in Sec.~\ref{sec:summary}.

\begin{table}[!tb]
\centering
\caption{
  Theoretical predictions~\cite{Bobeth:2013uxa} and experimental results for time-integrated branching fractions of \B meson decays to dilepton final states, $\overline{\cal B}(\decay{\Bds}{\ell^+\ell^-})$.
  Upper limits are at 90\% confidence level.
}
\label{tab:b2ll}
\vspace{2ex}
\begin{tabular}{ccccc}
  \hline
  Decay & Prediction & Measurement \\ 
  \hline
  \decay{\Bz}{\epem} & $(2.48 \pm 0.21) \times 10^{-15}$ & $<8.3 \times 10^{-8}$~\cite{Aaltonen:2009vr} \\
  \decay{\Bs}{\epem} & $(8.54 \pm 0.55) \times 10^{-14}$ & $<2.8 \times 10^{-7}$~\cite{Aaltonen:2009vr} \\
  \decay{\Bz}{\mumu} & $(1.06 \pm 0.09) \times 10^{-10}$ & $(3.9\,^{\,+1.6}_{\,-1.4}) \times 10^{-10}$~\cite{LHCb-PAPER-2014-049} \\
  \decay{\Bs}{\mumu} & $(3.65 \pm 0.23) \times 10^{-9}$ & $(2.8\,^{\,+0.7}_{\,-0.6}) \times 10^{-9}$~\cite{LHCb-PAPER-2014-049} \\
  \decay{\Bz}{\tautau} & $(2.22 \pm 0.19) \times 10^{-8}$ & $< 4.1 \times 10^{-3}$~\cite{Aubert:2005qw} \\
  \decay{\Bs}{\tautau} & $(7.73 \pm 0.49) \times 10^{-7}$ & No result$^{\dag}$ \\
  \hline
\end{tabular}
\\ $^{\dag}$ {\footnotesize A limit $\BR(\decay{\Bs}{\tautau}) < 5.0 \%$ has been derived~\cite{Grossman:1996qj} from ALEPH data~\cite{Buskulic:1994gj}.}
\end{table}

\subsection{Radiative decays}

As mentioned in Sec.~\ref{sec:intro}, measurements of the \decay{B}{X_s\gamma} branching fraction, performed by \babar\ and Belle~\cite{Lees:2012ym,Lees:2012ufa,Limosani:2009qg} are consistent with the SM expectation~\cite{Misiak:2006zs}.  
Therefore, the main focus in this area has switched to ratios of processes mediated by $b \to d\gamma$ and $b \to s\gamma$ transitions, \CP and isospin asymmetry measurements, and measurements of the polarisation of the emitted photon.

The ratio of branching fractions for $b\to d\gamma$ and $b\to s\gamma$ mediated decays determines the ratio of CKM matrix elements $\left| V_{td} / V_{ts} \right|^2$, and is therefore of interest to search for non-MFV BSM signatures.
The measurement of the inclusive $\BR(\decay{B}{X_d\gamma)}$ is very challenging for any experiment, but has nonetheless been performed by \babar~\cite{delAmoSanchez:2010ae}.
Perhaps more promising is the possibility to use exclusive decays, such as $\BR(\decay{\Bd}{\rho^0\gamma})/\BR(\decay{\Bd}{\Kstarz\gamma})$~\cite{Ali:2004hn,Ball:2006eu}.
Results have been presented by both \babar~\cite{Aubert:2008al} and Belle~\cite{Taniguchi:2008ty}, and give a precision on $\left| V_{td} / V_{ts} \right|$ of about 10\%.
Further reduction in the experimental uncertainty can be anticipated with results from LHCb, which has demonstrated its potential to reconstruct \decay{\Bd}{\Kstarz\gamma}~\cite{LHCb-PAPER-2012-019} by making the most precise determination of its \CP asymmetry.
Indeed, \CP asymmetries of both inclusive and exclusive radiative \bquark~hadron decays offer powerful null tests of the SM~\cite{Kagan:1998bh}, as do isospin asymmetries (\ie\ differences between charged and neutral \B meson decay rates)~\cite{Kagan:2001zk}.
All such measurements to date are consistent with the SM as shown in Table~\ref{tab:sgamma}.

\begin{table}[!tb]
  \centering
  \caption{
    Measurements of \CP and isospin asymmetries in \decay{\bquark}{\squark\gamma} transitions~\cite{HFAG}. 
    The value of $\mathcal{A}_{\CP}(\decay{B}{\Kstar\gamma})$ is the average for the \Bz decay which is much more precise than the value for the \Bp decay.  
    The value of $\mathcal{A}_{\CP}(\decay{B}{X_s\gamma})$ is dominated by a result from \babar~\cite{Lees:2014uoa}.
  }
  \label{tab:sgamma}
  \begin{tabular}{c|c|c}
     & \decay{B}{X_s\gamma} & \decay{B}{\Kstar\gamma} \\
     \hline
     $\mathcal{A}_{\CP}$   & $\phantom{-}0.015 \pm 0.020$ & $-0.002 \pm 0.015$ \\ 
     $\mathcal{A}_{\rm I}$ & $-0.01 \pm 0.06$  & $\phantom{-}0.012 \pm 0.051$ \\
  \end{tabular}
\end{table}

The branching fractions of inclusive and exclusive \decay{b}{s\gamma} decays are proportional at leading order to the photon dipole operator squared, $|C^{}_7|^{2} + |C^\prime_7|^{2}$, and are not sensitive to the handedness of the emitted photon. 
In the SM, photons produced in radiative \bquark~hadron decays are almost entirely left-handed due to the chiral nature of the charged current interaction. 
The right-handed component is suppressed by the ratio $C^\prime_7/C^{}_7 \sim m_s/m_b$, with the exact level of suppression being mode dependent due to QCD effects~\cite{Ball:2006eu,Grinstein:2004uu}. 
In many models that extend the SM, the virtual particles responsible for mediating the decay have no preferred left- or right-handed coupling and the photon can be produced with a significantly lower degree of polarisation. 
Well-known examples include supersymmetric models beyond MFV, left-right symmetric models, leptoquarks and models with additional gauge-bosons.

Several methods to measure the photon polarisation have been proposed.
One of the most promising exploits the interference between \Bz and \Bzb decays when the hadronic system in the \decay{B}{f\gamma} decay is accessible to both.
In such a case, the decay time dependent asymmetry can be written
\begin{equation}
  \mathcal{A}_{\CP}[ B \to f \gamma ](t) = S_{f\gamma} \sin ( \Delta m t ) - C_{f\gamma} \cos ( \Delta m t ) ~,
\end{equation} 
where $\Delta m$ is the mass difference in the $\Bd$--$\Bdb$ system.
The \CP asymmetry in decay, $C_{f\gamma}$, has the same sensitivity as that measured with decay-time integrated methods, but the coefficient of the sinusoidal oscillation can be written~\cite{Atwood:1997zr,Atwood:2004jj}
\begin{equation}
  S_{f\gamma} = \chi_f \sin(2\psi) \sin(2\beta) ~,
\end{equation} 
where $\chi_f$ is the $C$ eigenvalue of the hadronic system, $\tan\psi$ gives the magnitude of the ratio of right- and left-handed amplitudes and $\beta \equiv \arg\left[ - \frac{V_{cd}^{}V_{cb}^*}{V_{td}^{}V_{tb}^*} \right]$ is the angle of the CKM Unitarity Triangle, which is measured to be $\sin(2\beta) = 0.682 \pm 0.019$~\cite{HFAG}, assuming no NP in $\Bd$--$\Bdb$ oscillations.
Measurements of $S_{f\gamma}$ can therefore be interpreted in terms of $\sin(2\psi)$, and thus in terms of $C^\prime_7/C^{}_7$.

The coefficient $S_{K^{*}\gamma}$ has been measured by \babar and Belle using \decay{\Bz}{\Kstarz\gamma} decays where \decay{\Kstarz}{\KS\piz}, giving~\cite{HFAG,Aubert:2008gy,Ushiroda:2006fi}
\begin{equation} 
  S_{\Kstar\gamma} = -0.16\pm 0.22
\end{equation} 
which is consistent with the SM prediction of $-0.02$~\cite{Ball:2006eu}. 
Measurements of similar coefficients in different final states are somewhat less precise, though the results for \decay{\Bz}{\KS\rho^0\gamma} are competitive~\cite{Li:2008qma,Akar:2014hda}.
Significant improvement in the sensitivity to right-handed currents is a key goal of current and future experiments.

The \decay{\Bz}{\KS\piz\gamma} decay is highly challenging to reconstruct at the LHC, but the production of all \bquark~hadron species opens alternative possibilities.
Suggestions to measure the photon polarisation using \decay{\Lb}{\Lz^{(*)}\gamma} decays~\cite{Legger:2006cq, Hiller:2007ur} at the LHC have proved experimentally difficult due to the small polarisation of \Lb baryons produced at the LHC~\cite{LHCb-PAPER-2012-057}.
However, the \decay{\Bs}{\phi\gamma} decay appears attractive.
Although the small SM value of $\beta_s \equiv - \arg\left[ - \frac{V_{cs}^{}V_{cb}^*}{V_{ts}^{}V_{tb}^*} \right]$ suppresses the $S_{f\gamma}$ coefficient, the non-zero value of the width difference $\Delta \Gamma_s$ in the $\Bs$--$\Bsb$ system results in sensitivity to the photon polarisation through the effective lifetime, or equivalently the $A^{\Delta\Gamma}$ parameter, of \decay{\Bs}{\phi\gamma} decays~\cite{Muheim:2008vu}.
An experimental advantage of such an analysis is that it does not require flavour tagging.
LHCb has previously shown that it can reconstruct large yields of \decay{\Bs}{\phi\gamma} decays~\cite{LHCb-PAPER-2012-019}, and its first results on the effective lifetime are keenly anticipated.

\begin{figure}[!tb]
\centering
\includegraphics[width=0.8\linewidth]{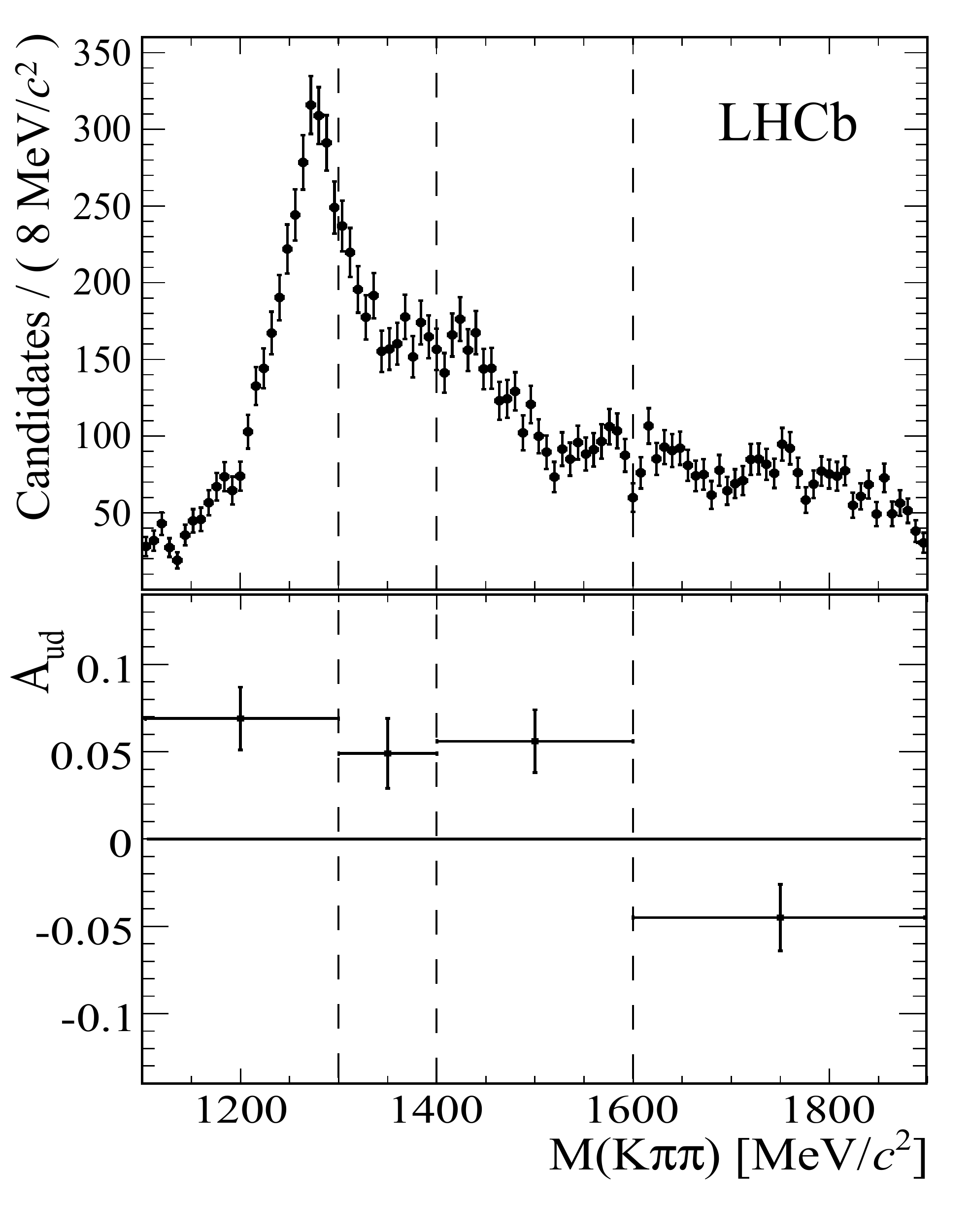}
\caption{
  Top: background subtracted $\Kp\pim\pip$ invariant mass distribution in \decay{\Bp}{\Kp\pim\pip\gamma} decays. 
  Bottom: up-down asymmetry ${\rm A_{ud}}$ in bins of $M(\Kp\pim\pip)$. 
  Modified from Ref.~\cite{LHCb-PAPER-2014-001}. 
  \label{fig:updwon}
} 
\end{figure}

Another way to probe $C^\prime_7/C^{}_7$  is through the photon direction with respect to the plane defined by the $\pip\pim$ system in \decay{\Bp}{\Kp\pim\pip\gamma} decays~\cite{Gronau:2001ng,Gronau:2002rz,Kou:2010kn}. 
The so-called up-down asymmetry of the photon with respect to this plane is proportional to the photon polarisation.
The constant of proportionality suffers large hadronic uncertainties, that can, however, be controlled to some extent from data on \decay{\Bp}{\jpsi\Kp\pip\pim} decays and other information concerning the $K\pi\pi$ system.
LHCb has performed a first measurement of this up-down asymmetry using its full Run~I dataset~\cite{LHCb-PAPER-2014-001}. 
As shown in Fig.~\ref{fig:updwon}, the data are split into four regions of $\Kp\pim\pip$ mass ($M(\Kp\pim\pip)$) defined by the known $\Kp\pim\pip$ resonances. 
Whilst non-zero photon polarisation is observed at the level of $5.2\sigma$, when combining the four regions of $M(\Kp\pim\pip)$, a deeper understanding of the structure of the $\Kp\pim\pip$ system is needed to determine $C^\prime_7/C^{}_7$ with this approach.

The \decay{\Bz}{\Kstarz\ell^{+}\ell^{-}} decays can also be used to determine $C^\prime_7/C^{}_7$ at low dilepton invariant masses, where virtual photon contributions are expected to dominate. 
Study of the angular distributions allows the determination of the parameter $A_{\rm T}^{(2)}$ defined in Eq.~(\ref{eq:AT2}).
At the limit $\qsq \to 0$ this quantity is directly sensitive to $C^\prime_7/C^{}_7$ (considering complex valued Wilson coefficients, another observable $A_{\rm T}^{\rm Im}$ probes the relative phase between $C^\prime_7$ and $C^{}_7$).
LHCb has studied $\Bz \to \Kstarz \epem$ decays in the low $\qsq$ region~\cite{LHCb-PAPER-2013-005,LHCb-PAPER-2014-066}, obtaining
\begin{equation}
  A_{\rm T}^{(2)} = -0.23 \pm 0.23 \pm 0.05
\end{equation}
for $0.002 < \qsq < 1.120 \gev^2/c^4$ from the Run~I dataset.

\subsection{Semileptonic \decay{b}{s\ell^+\ell^-} decays} 
\label{sec:expt:kstarll}

The LHC data has led to a wealth of results in semileptonic $b \to s \ell^+\ell^-$ decays.  
In the following subsections results on branching fractions and rate asymmetries are first discussed, followed by considerations on analyses of angular distributions.

\subsubsection*{Branching fractions of semileptonic \decay{b}{s\ell^+\ell^-} decays} 

In contrast to the case of $b \to s\gamma$ decays, most of the experimental work on the $b \to s\ell^+\ell^-$ process has been with exclusive final states.
Results on the inclusive decay~\cite{Lees:2013nxa,Sato:2014pjr} do not yet reach high precision, but will be an important topic in the future.
The LHC data have, however, led to a large increase in the yields of certain $b \to s\ell^+\ell^-$ decays, in particular those with a dimuon pair in the final state. 
This has led to increasingly precise determinations of the branching fractions of the \decay{B}{K\mumu}, \decay{B}{\Kstar\mumu} and \decay{\Bs}{\phi\mumu} decays~\cite{LHCb-PAPER-2014-006,LHCb-PAPER-2013-019,Chatrchyan:2013cda,LHCb-PAPER-2013-017}, as well as of the semileptonic \bquark~baryon decay \decay{\Lb}{\Lz\mumu}~\cite{Aaltonen:2011qs,LHCb-PAPER-2013-025}. 
The theory input for the baryon decays~\cite{Boer:2014kda,Wang:2008sm}, in particular knowledge of the form-factors, is less well advanced than for the mesons, so they are not discussed further in this review.  

The experimental results on the heavy-to-light branching fractions are now much more precise than the corresponding theoretical predictions, with further improved measurements anticipated in the coming years. 
This is illustrated in Table~\ref{tab:BKll}, where measurements of $\BF(\decay{\Bp}{\Kp\mumu})$ are compared to theory predictions in the low and high \qsq regions. 
SM predictions for the branching fractions are sensitive to hadronic uncertainties in the form factors, which typically lead to uncertainties of $\mathcal{O}(30\%)$ on the SM predictions. 
Progress from lattice QCD has improved the precision in the high \qsq region above the $\psi(3770)$ resonance, but the sensitivity to BSM physics still remains limited by the uncertainty on the SM predictions. 

\begin{table}[!tb]
  \centering
  \caption{Branching fraction of the \decay{\Bp}{\Kp\mumu} decay in selected $\qsq$ bins, $[\qsq_{\rm min}, \qsq_{\rm max}]$ (in $\gev^{2}/c^4$), from LHCb~\cite{LHCb-PAPER-2012-024} compared to SM predictions using light-cone-sum-rule (LCSR)~\cite{Khodjamirian:2012rm} and lattice~\cite{Bouchard:2013eph} calculations of form factors.
  The first uncertainty on the experimental results is statistical and the second is systematic.}
  \label{tab:BKll}
  \vspace{1ex}
\resizebox{\textwidth}{!}{
  \begin{tabular}{l|c|c|c}
    &  LHCb &  SM (LCSR)  & SM (lattice)\\ \hline
$[1,6]$ & $(1.21 \pm 0.09 \pm 0.07) \times 10^{-7}$ & $(1.75\,^{+0.60}_{-0.29})  \times 10^{-7}$ &  $(1.81 \pm 0.61) \times 10^{-7}$\\
$[16,18]$ & $(0.35 \pm 0.04\pm 0.02) \times 10^{-7}$ & $(0.33\,^{+0.19}_{-0.09})  \times 10^{-7}$ & $(0.39 \pm 0.04) \times 10^{-7}$ \\
  \end{tabular}
}
\end{table}

In order to increase sensitivity to BSM physics, it is useful to study observables in which the effects of form factor uncertainties are reduced. 
Two such quantities are the \CP asymmetry between \B and \Bb decays, $\mathcal{A}_{\CP}$, and the isospin asymmetry, $\mathcal{A}_{\rm I}$ between \Bp and \Bz decays.
In the SM, the \CP asymmetries of \decay{\B}{\Kstar\mumu} and \decay{\B}{K\mumu} decays are tiny, $\mathcal{O}(10^{-3})$, due to the small numerical size of the product of CKM elements $V_{ub}^{}V_{us}^{*}$ compared to $V_{tb}^{}V_{ts}^{*}$. 
Extensions of the SM can provide new sources of \CP violation, and thus $\mathcal{A}_{\CP}$ constitutes a null test of the SM, where any visible direct \CP violation would be evidence for BSM physics.  
The latest results, all consistent with the SM expectation of close to zero, are summarised in Table~\ref{tab:smumu}.

The decays appearing in the isospin asymmetry only differ by the flavour of the spectator quark in the \B meson (\uquark~quark for the \Bp meson and \dquark~quark for the \Bz).
In the framework of the effective Hamiltonian, $\mathcal{A}_{\rm I}$ differs from zero due to isospin breaking effects in the form factors, from annihilation and exchange amplitudes, and from spectator scattering, where a virtual photon is emitted. 
The isospin asymmetry is expected to be $\approx -1\%$ at large \qsq in the SM~\cite{Beylich:2011aq} and for $B \to \Kstar$ decays to grow to $\approx +10\%$ as $\qsq$ tends to zero~\cite{Feldmann:2002iw,Kagan:2001zk}.  
Using its full Run~I dataset, LHCb found $\mathcal{A}_{\rm I}$ in $B \to K$ and $B \to \Kstar$ decays to be consistent with zero across the full \qsq window~\cite{LHCb-PAPER-2014-006} (see Table~\ref{tab:smumu}).

\begin{table}[!tb]
  \centering
  \caption{
    Measurements of \CP and isospin asymmetries in \decay{\B}{K^{(*)}\mu^+\mu^-} transitions at low and high values of \qsq~\cite{LHCb-PAPER-2014-032,LHCb-PAPER-2014-006}. 
    Units of $\gev^2/c^4$ for \qsq are implied.
    Only results from LHCb are included since earlier, less precise, measurements used a different \qsq binning scheme.
  }
  \label{tab:smumu}
  \vspace{1ex}
\resizebox{\textwidth}{!}{
  \begin{tabular}{c|cc|cc}
    & \multicolumn{2}{c}{\decay{B}{K\mu^+\mu^-}} 
    & \multicolumn{2}{c}{\decay{B}{\Kstar\mu^+\mu^-}} \\
    \qsq range
    & $1.1$--$6.0$ & $15.0$--$22.0$
    & $1.1$--$6.0$ & $15.0$--$19.0$ \\
    \hline
    $\mathcal{A}_{\CP}$   
    & $ 0.004 \pm 0.028$ & $-0.005 \pm 0.030$
    & $-0.094 \pm 0.047$ & $-0.074 \pm 0.044$ \\
    $\mathcal{A}_{\rm I}$ 
    & $-0.10 \,^{+0.08}_{-0.09} \pm 0.02$ & $-0.09 \pm 0.08 \pm 0.02$
    & $ 0.00 \,^{+0.12}_{-0.10} \pm 0.02$ & $ 0.06 \,^{+0.10}_{-0.09} \pm 0.02$ \\
  \end{tabular}
}
\end{table}

The ratio of rates for $b \to d\ell^{+}\ell^{-}$ and $b \to s\ell^{+}\ell^{-}$ processes is sensitive to $\left|V_{td}/V_{ts}\right|^{2}$.  
The observation by LHCb of the \decay{\Bp}{\pip\mumu} decay gives~\cite{LHCb-PAPER-2012-020}
\begin{equation}
\begin{split}
  {\cal B}(\decay{\Bp}{\pip\mumu}) &= (2.3 \pm 0.6 \;({\rm stat}) \pm 0.1 \;({\rm syst})) \times 10^{-8} ~, \\
  \left|V_{td}/V_{ts}\right| &= 0.266 \pm 0.035 \;({\rm stat}) \pm 0.003 \;({\rm syst}) ~,
\end{split} 
\end{equation}
where the uncertainty on $\left|V_{td}/V_{ts}\right|$ due to knowledge of the form factors, estimated to be 5.1\%, is not included in the result.
Further improvements, and observations of more $b \to d\ell^{+}\ell^{-}$ decay modes, are anticipated in the coming years.

\subsubsection*{Angular analyses of \decay{b}{s\ell^+\ell^-} decays}

An angular analysis of $B \to K \ell^+ \ell^-$ decays provides a simple null test of the SM~\cite{Bobeth:2007dw}. 
The angular distribution can be described by a single angle $\theta_{\ell}$
\begin{align}
  \label{eq:babar}
  \frac{1}{\Gamma} \frac{\deriv\Gamma (B \to K \ell^+ \ell^-)}{\deriv\cos\theta_{\ell}}
    = \frac{3}{4} (1 - F^{}_{\rm H}) (1 - \cos^2\theta_{\ell}) 
    + \frac{1}{2} F^{}_{\rm H} + A^{}_{\rm FB} \cos\theta_{\ell} ,
\end{align} 
with a constant term, $F^{}_{\rm H}/2$, and a forward-backward asymmetry, $A^{}_{\rm FB}$, linear in $\cos \theta_{\ell}$. 
Both $F^{}_{\rm H}$ and $A^{}_{\rm FB}$ are small within the SM, for $\ell = e, \mu$, and therefore can signal the presence of BSM physics.
In particular these terms are sensitive to contributions from new scalar, pseudoscalar and tensor operators.
LHCb has made precise measurements of these parameters in the decay \decay{\Bp}{\Kp\mumu}~\cite{LHCB-PAPER-2014-007}. 
The measurements are consistent with $A^{}_{\rm FB} = 0$ and $F^{}_{\rm H} \approx 0$, expected in the SM.

The angular distribution of the \decay{\Bz}{\Kstarz\mumu} decay, with \decay{\Kstarz}{\Kp\pim}, is more complicated, and can be described by three angles: $\theta_{\ell}$, which is defined by the direction of the \mup (\mun) with respect to the \Bz (\Bzb) in the dimuon rest frame; $\theta_{K}$, which is  defined by the direction of the kaon with respect to the \Bz (\Bzb) in the \Kstarz (\Kstarzb) rest frame; and $\phi$, the angle between the plane containing the \mup and \mun and the plane containing the kaon and pion. 
The differential decay rates in terms of these angles and the dimuon invariant mass squared, for \Bz and \Bzb decays, are given by 
\begin{equation}
\begin{split}
  \frac{\deriv^{4}\Gamma[\Bz \to\Kstarz \mumu]}{\deriv\cos\theta_{\ell}\,\deriv\cos\theta_{K}\,\deriv\phi\,\deriv\qsq} & = \frac{9}{32\pi} \sum_{i} \bar{J}_{i}(\qsq) f_{i}(\cos\theta_{\ell},\cos\theta_{K},\phi)~, \\
  \frac{\deriv^{4}\Gamma[\Bzb\to\Kstarzb\mumu]}{\deriv\cos\theta_{\ell}\,\deriv\cos\theta_{K}\,\deriv\phi\,\deriv\qsq} & = \frac{9}{32\pi} \sum_{i} {J}_{i}(\qsq) f_{i}(\cos\theta_{\ell},\cos\theta_{K},\phi)~. \\
\end{split}
  \label{eq:kstarmumu-angular}
\end{equation} 
Here, the $f_{i}(\cos\theta_{\ell},\cos\theta_{K},\phi)$ originate from spherical harmonics and the $J_i$ and $\bar{J}_i$ are bilinear combinations of \Kstarz decay amplitudes ($A_{\parallel}^{\rm L,R}$, $A_{\perp}^{\rm L,R}$ and $A_{0}^{\rm L,R}$)~\cite{Kruger:1999xa}.
The \CP averaged observables,
\begin{align}
S_i = (J_i + \bar{J}_i) \Bigg/ \frac{\deriv\left( \Gamma + \overline{\Gamma} \right)}{\deriv\qsq} 
\end{align}
depend on the underlying short distance contributions from $C^{}_7 \pm C^\prime_7$, $C^{}_9 \pm C^\prime_9$ and $C^{}_{10} \pm C^\prime_{10}$, for \CP odd (even) components.
They can be related to the observables discussed in Sec.~\ref{sec:symmetry}, for example by $S^{}_{3} = (1 - F^{}_{\rm L}) A^{(2)}_{\rm T}/2$, $S^{}_{4,5} = \sqrt{F^{}_{\rm L} ( 1- F^{}_{\rm L} )} P^\prime_{4,5}$.
In addition, \CP violating observables ($\propto J_i - \bar{J}_i$) can be obtained from the angular distributions, including several with high BSM sensitivity~\cite{Kruger:1999xa,Bobeth:2008ij,Egede:2010zc,Alok:2011gv}.

\begin{figure}[!tb]
\centering
\includegraphics[width=0.48\textwidth]{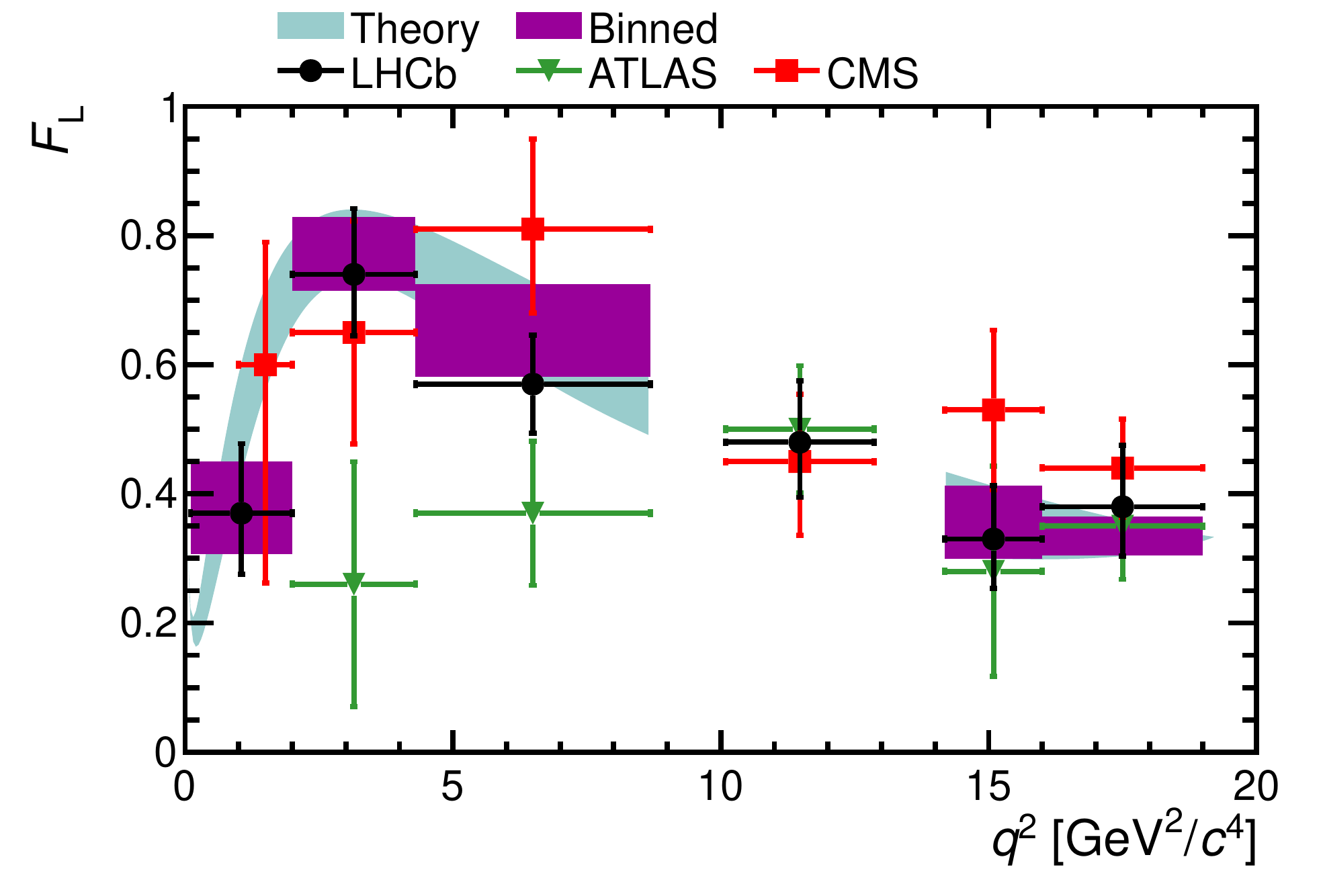} 
\includegraphics[width=0.48\textwidth]{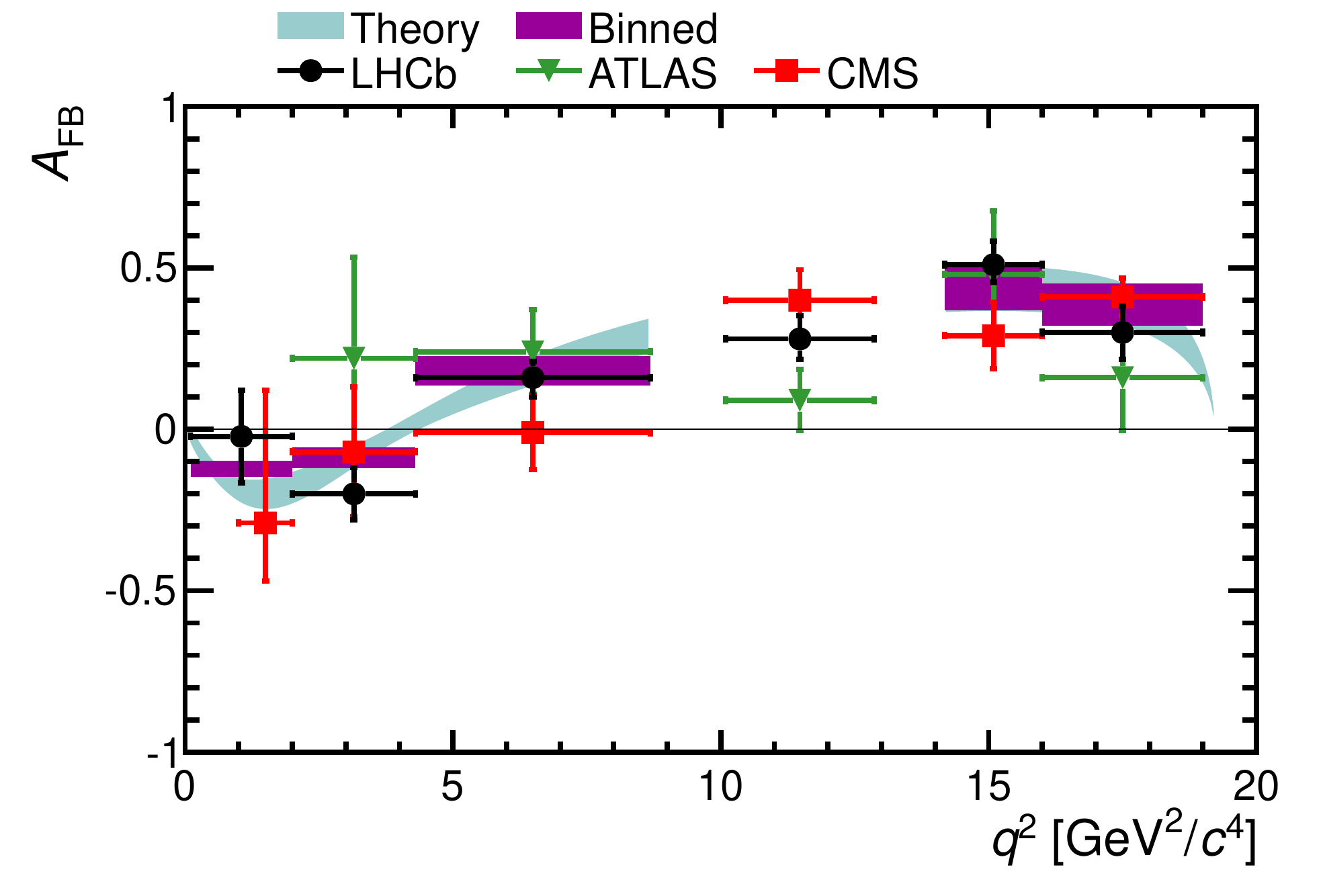}
\caption{
  (Left) longitudinal polarisation fraction, $F_{\rm L}$, of \Kstarz mesons produced in \decay{\Bz}{\Kstarz\mumu} decays.
  (Right) dimuon system forward-backward asymmetry, $A_{\rm FB}$. 
  Results from ATLAS~\cite{ATLAS:KstarMuMu}, CMS~\cite{Chatrchyan:2013cda} and LHCb~\cite{LHCb-PAPER-2013-019} are included. 
  The data are overlaid with an SM prediction~\cite{Bobeth:2011gi}.} 
\label{fig:angular}
\end{figure}

Prior to data taking at the LHC, with the relatively modest samples of \decay{\Bz}{\Kstarz\mumu} decays that were available, it was not possible to determine all of the terms of Eq.~(\ref{eq:kstarmumu-angular}). 
Instead, partial angular analyses of the decay were performed using single angle projections.
ATLAS, CMS and LHCb have also performed similar analyses~\cite{ATLAS:KstarMuMu,Chatrchyan:2013cda,LHCb-PAPER-2013-019} giving sensitivity to $F_{\rm L}$, $A_{\rm FB}$ (see Fig.~\ref{fig:angular}) and, in the case of LHCb, $A_{\rm T}^{(2)}$.
Each of these observables is consistent with the SM expectation. 

\begin{figure}[!tb]
\centering
\includegraphics[width=0.48\textwidth]{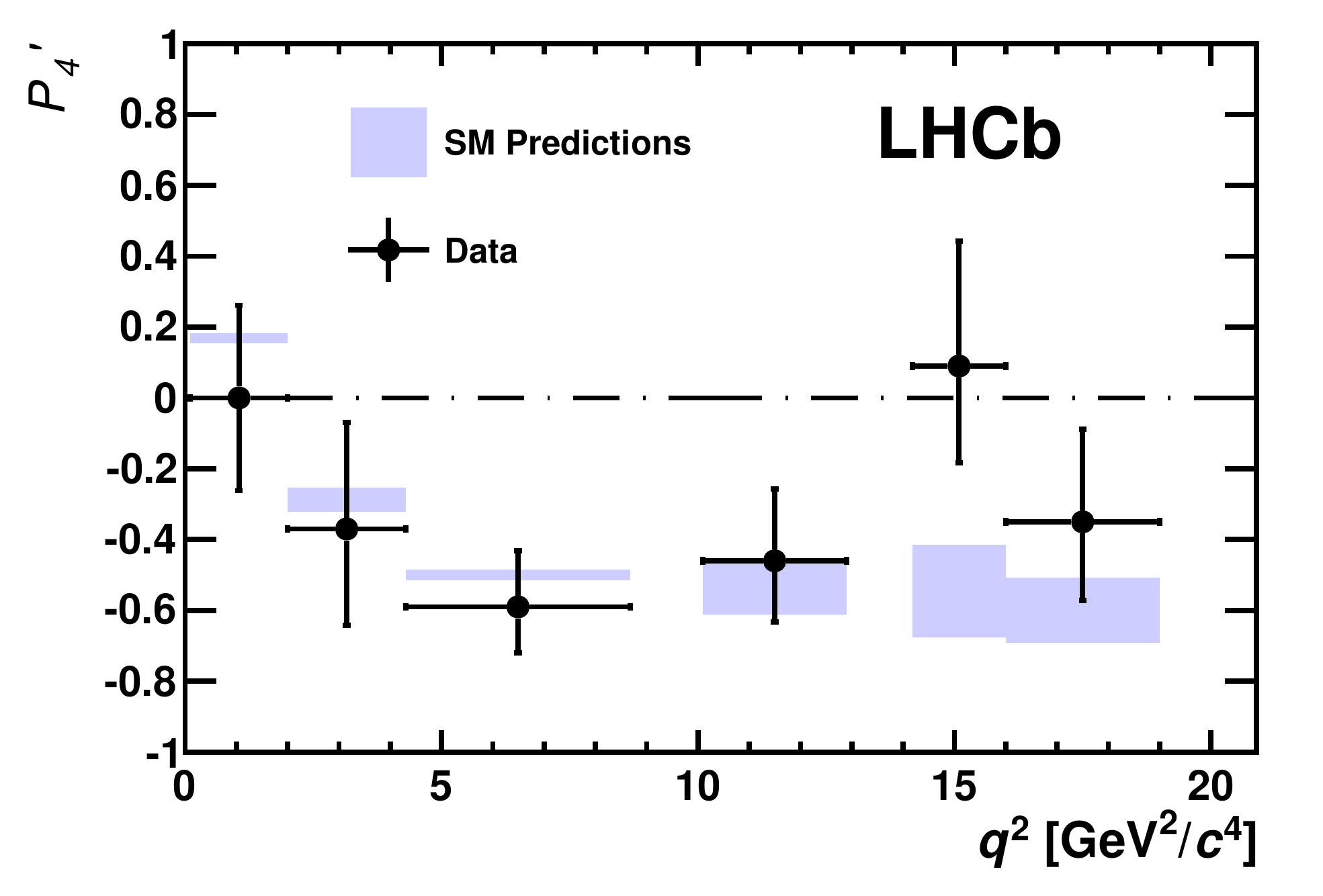} 
\includegraphics[width=0.48\textwidth]{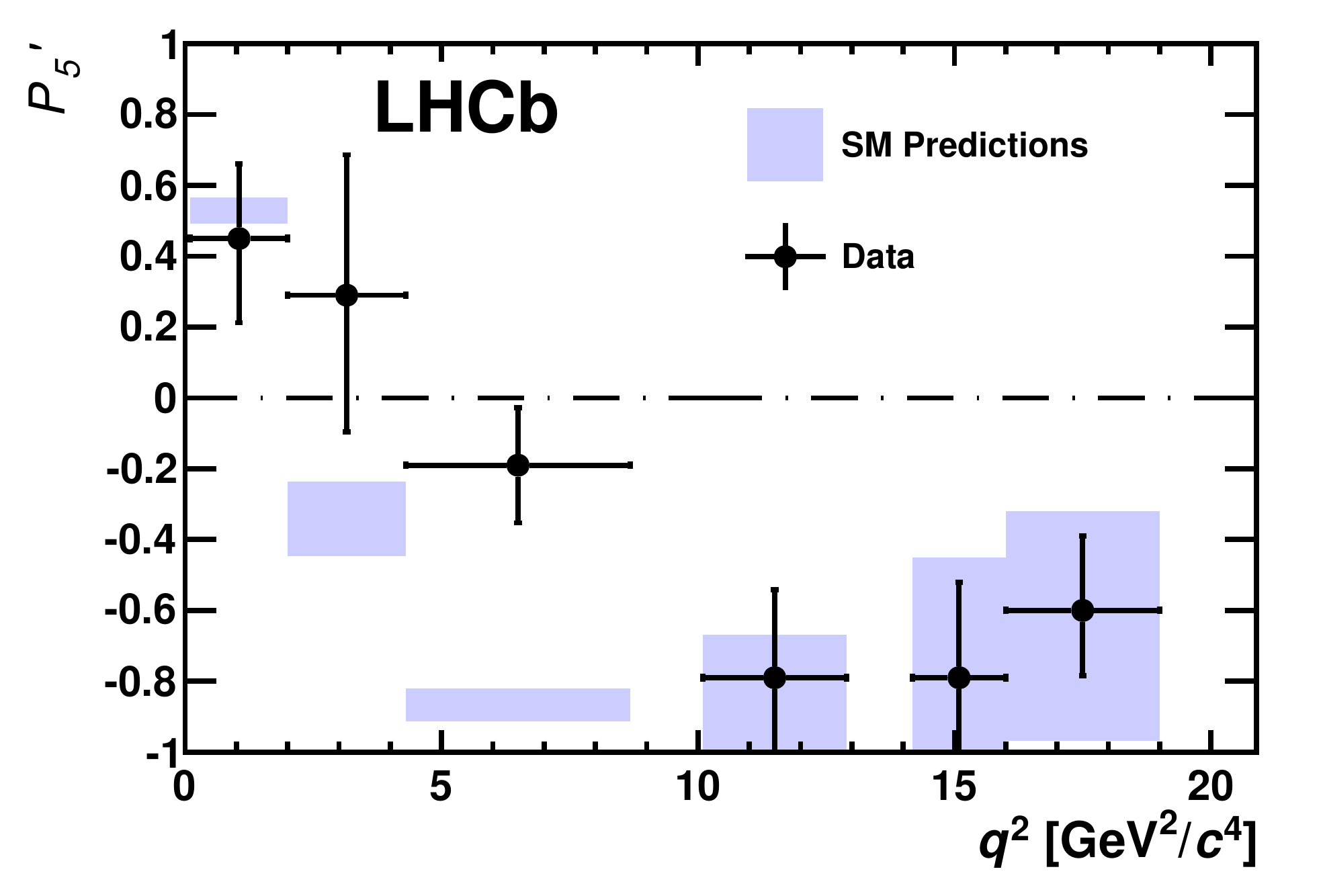}
\caption{Observables $P^\prime_4$ and $P^\prime_5$, measured by LHCb~\cite{LHCb-PAPER-2013-037} in \decay{\Bz}{\Kstarz\mumu} decays. The data are overlaid with an SM prediction~\cite{Descotes-Genon:2013vna}.} 
\label{fig:p5prime}
\end{figure}

LHCb has also measured two of the ``optimised'' observables discussed in Sec.~\ref{sec:symmetry}, $P^\prime_{4}$ and $P^\prime_{5}$~\cite{LHCb-PAPER-2013-037}. 
In the low \qsq region, there is a large local discrepancy between the data for $P'_{5}$ and the SM expectation at the level of $3.7\sigma$, as shown in Fig.~\ref{fig:p5prime}. 
This is discussed further in Sec.~\ref{sec:interpretation}.

It is expected that full angular analyses of the \decay{\Bz}{\Kstarz\mumu} decay should be possible with the full Run~I datasets of the LHC experiments. 
As the analyses become more precise, and more complex, it will also be important to account for the contribution from $K\pi$ S-wave under the \Kstarz peak~\cite{Becirevic:2012dp,Matias:2012qz,Blake:2012mb}.
It will also be possible to determine \CP asymmetries for each of the angular terms; these can then be used to constrain the imaginary parts of the Wilson coefficients. 
Other \decay{\B}{V\ell^+\ell^-} decays, such as \decay{\Bs}{\phi\mumu}, will provide additional constraints.

\subsection{Non-universal lepton couplings} 
\label{sec:expt:luv}

In the SM, with the notable exception of the Higgs boson, particles couple equally to the different flavours of lepton. 
The ratio of decay rates  
\begin{equation}
  R_{H} \equiv  \frac{\Gamma[\decay{\B}{H\mumu}]}{\Gamma[\decay{\B}{H\ep\en}]}
\end{equation}
where $H = K, \Kstar, X_s,$ \etc, is therefore expected to differ from unity only due to tiny Higgs penguin contributions and phase space differences~\cite{Hiller:2003js}. 
Using the full Run~I dataset, LHCb measured in the \qsq range $1 < \qsq < 6\gev^{2}/c^{4}$~\cite{LHCb-PAPER-2014-024}
\begin{equation}
  \label{eq:expt:rk}
  R_{K}[1,6] = 0.745 \,^{+0.090}_{-0.074} \;({\rm stat}) \pm 0.036 \;({\rm syst}) \, ,
\end{equation} 
which differs from the SM expectation of $R_{K} = 1.0003 \pm 0.0001$~\cite{Bobeth:2007dw} by $2.6\sigma$. 
Although not yet at the level of significance that qualifies as ``evidence'', this result has prompted theoretical speculation concerning possible sources of lepton non-universality, as discussed further in Sec.~\ref{sec:interpretation}.
Since results from \babar\ on \decay{\B}{D^{(*)}\tau\nu} decays~\cite{Lees:2012xj,Lees:2013uzd} also hint at violation of universality, this is a highly topical area.

\subsection{Null tests}

Null tests of the SM, \ie\ searches for signals that are absent or vanishingly small in the SM, are valuable for several reasons.
Observation of such a process would not only provide a smoking gun signature of BSM physics, but would also indicate in what way the NP should be accommodated within the operator basis discussed in Sec.~\ref{sec:MIA}.
In the absence of signals, limits can be placed on the contributions of the additional operators which, when sufficiently stringent, justify the use of a restricted set of operators in model-independent analyses.  

The tests of lepton universality discussed above fall into this category of null tests.
It is also important to explore the possibility of lepton flavour violation (LFV), and lepton number violation (LNV) in rare \bquark~hadron decays. 
The observation of neutrino oscillation demonstrates that lepton flavour is not an exact symmetry of nature, but if the SM is minimally expanded to allow neutrino mass, rates of processes with charged LFV remain unobservably small.
Charged LFV, and LNV, does however arise in many BSM theories.
While strong limits exist from searches for rare muon and tau decays (for LFV)~\cite{Marciano:2008zz} and for neutrinoless double beta decay (for LNV)~\cite{Avignone:2007fu}, there are models which respect those limits but nonetheless produce observable signatures in \bquark~hadron decays. 

One of the most powerful of such LFV searches is for the \decay{\Bds}{e^\pm\mu^\mp} decay.
The experimental limits have been improved by LHCb to the level of $\lesssim 10^{-8}$~\cite{LHCb-PAPER-2013-030}.
The limit on the \decay{\Bd}{e^\pm\mu^\mp} decay is four orders of magnitude more stringent than those on \decay{\Bd}{e^\pm\tau^\mp} and \decay{\Bd}{\mu^\pm\tau^\mp}~\cite{Aubert:2008cu}.
For semileptonic decays, the limits on the \decay{\Bp}{\pip e^\pm\mu^\mp} and \decay{\Bp}{\Kp e^\pm\mu^\mp} decays, at the level of $10^{-7}$~\cite{Aubert:2006vb,Aubert:2007mm}, are similarly much more stringent than those on decays involving $\tau$ leptons~\cite{Lees:2012zz}.
Since the operators can be written as being independent for each pair of leptons, it is important to improve all limits.
These LFV semileptonic decay modes have not yet been investigated at the LHC.

Several searches have been performed for LNV in \bquark~hadron decays into final states containing a pair of same sign leptons, \decay{\Bp}{M^-\ell^+\ell^{\prime +}}.
The strongest limits, from LHCb on $\BR(\decay{\Bp}{\pim\mup\mup})$, are at the level of $10^{-9}$~\cite{LHCb-PAPER-2013-064}.
These are complemented by limits on numerous modes with different hadronic systems $M^- = \pi^-, \rho^-, \Km, \Kstarm, \Dm, \Dstarm, \Dsm, ...$ and in which one or both leptons may be an electron, all at the level of $10^{-6}$ or below~\cite{LHCb-PAPER-2011-038,Seon:2011ni,Liventsev:2013zz,BABAR:2012aa,Lees:2013gdj}.
Limits also exist at the $10^{-6}$ level~\cite{BABAR:2011ac} on several \bquark~hadron decays that violate both baryon and lepton number, but these have not yet been explored at the LHC.

The most recent LHCb analysis of \decay{\Bp}{\pim\mup\mup}~\cite{LHCb-PAPER-2013-064} sets limits on the branching fraction as functions of the mass and decay time of the $\pim\mup$ pair.
These results are of interest to probe models where the decay is mediated by the on-shell production of a Majorana neutrino, which could be long-lived.
Similar experimental techniques can be exploited to search for long-lived particles ($X$) in \decay{B}{K X} with \decay{X}{\ell^+\ell^-} and similar decays. 
Such signatures are predicted in a range of theories, that are generically referred to as ``dark sector'' models~\cite{Essig:2013lka}. 
Although these decays have not yet been investigated at the LHC, the possibility to search for particles that travel ${\cal O}(1\,{\rm m})$ before decaying makes these probes complementary to other searches for new light resonances.

\section{Interpretation}
\label{sec:interpretation}

\subsection{Wilson coefficient fits}
\label{sec:inter:wilson}

The observation of the \decay{\Bs}{\mumu} decay, described in Sec.~\ref{sec:expt:dimuon}, puts strong constraints on scalar and pseudo-scalar operators ($\mathcal{O}^{(\prime)}_{\rm S}$ and $\mathcal{O}^{(\prime)}_{\rm P}$).  
In BSM models, the \decay{\Bs}{\ell^+\ell^-} branching fraction is enhanced or suppressed by the ratio
\begin{equation} 
\frac{ {\cal{B}}(\Bsb  \to  \ell^+ \ell^-)}{{\cal{B}}(\Bsb \to  \ell^+ \ell^-)_{\rm SM} } =
   |1- 0.24 (C_{10}^{\ell\,{\rm NP}} -C_{10}^{ \ell \prime})-y_\ell (C_{\rm P}^{\ell}- C_{\rm P}^{\ell \prime}) |^2
    + | y_\ell  (C_{\rm S}^{\ell } - C_{\rm S}^{\ell \prime})|^2  ~,
\end{equation}
where $y_\mu = 7.7$ and $y_e =(m_\mu/m_e) y_\mu=1.6 \times 10^3$. 
At $1\sigma$, the current experimental measurements imply~\cite{Bobeth:2012vn}
\begin{equation}
  \label{eq;Sboundmu}
   | C^{\mu}_{\rm P} -C_{\rm P}^{\mu \prime}| \lesssim 0.3 \, ~~\text{and}~~ |C^{\mu}_{\rm S} - C_{\rm S}^{\mu \prime}| \lesssim 0.1 ~.
\end{equation}
The constraints for dielectron decays are somewhat weaker,
\begin{equation}
  \label{eq;Sbounde}
 |C_{\rm S}^e- C_{\rm S}^{e \prime}|^2 + |C_{\rm P}^e - C_{\rm P}^{e \prime}|^2  \lesssim 1.3   \, ,
\end{equation}
and substantial room still exists for ditau decays.
Barring fortuitous cancellations that can happen if $C^{}_{\rm S} = C_{\rm S}^{\prime}$ and $C^{}_{\rm P} = C_{\rm P}^{\prime}$, visible effects from operators $\mathcal{O}_{\rm S}^{(\prime)}$ and $\mathcal{O}_{\rm P}^{(\prime)}$ to dimuon and dielectron decays can be neglected.
Large accidental cancellations are also excluded by $B \to K \ell^+ \ell^-$ decays, which constrain the combinations $|C^{}_{\rm S} + C^\prime_{\rm S}|$ and $|C^{}_{\rm P} + C^\prime_{\rm P}|$.
Contributions from tensor operators, are also constrained by \decay{B}{K\ell^+\ell^-} decays. 
In particular, the small size of the $F_{\rm H}$ term in the \decay{\Bp}{\Kp\mumu} angular distribution~\cite{Hiller:2014yaa}, leads to a bound of
\begin{align} 
  \label{eq;Tboundmu}
   |C_{\rm T}|^2 + |C_{\rm T5}|^2 & \lesssim 0.5 \, .
 \end{align}
Therefore, it is appropriate to focus on the operator basis of Eq.~(\ref{eq:operators}), and consider what the current measurements tell us about BSM contributions. 

\begin{figure}[!tb]
\centering
\includegraphics[width=\linewidth]{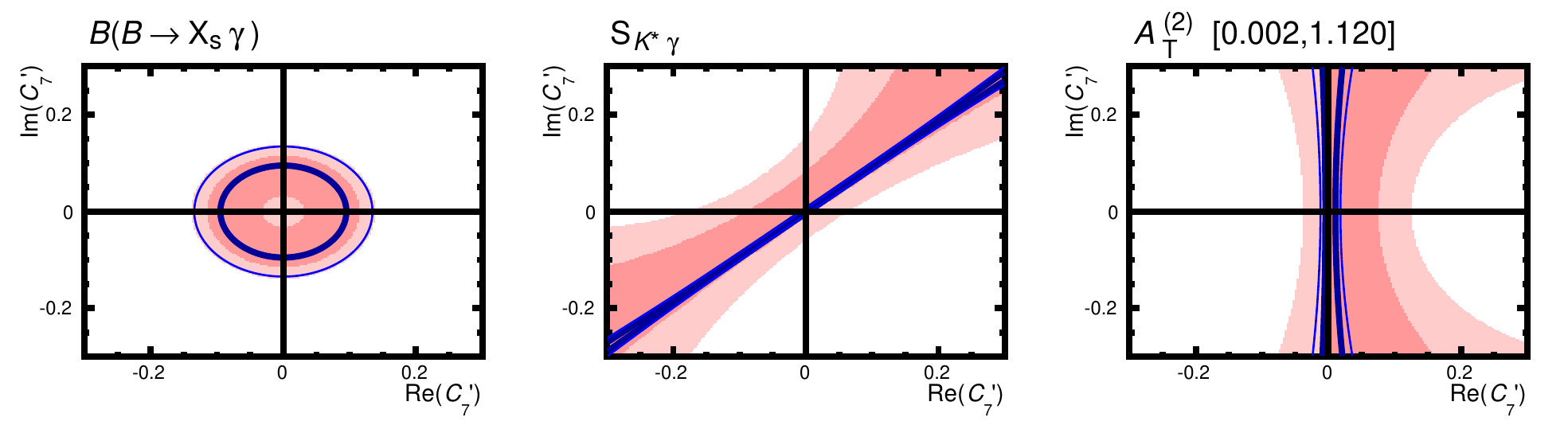}
\caption{
  Dependence of the observables $\BF(B \to X_{s} \gamma)$,  $S_{\Kstar\gamma}$ and $A_{\rm T}^{(2)}$ in the \qsq range $0.002 < \qsq < 1.120 \gev^{2}/c^{4}$ on ${\rm Re}(C_{7}^{\prime})$ and ${\rm Im}(C_{7}^{\prime})$, computed using the EOS flavour tool~\cite{Bobeth:2010wg} with all other Wilson coefficients fixed to their SM expectations.
  The contours (red) indicate the 68\% and 90\%  intervals on experimental determinations of the observables. 
  The lines indicate the theoretical uncertainty on each observable.
  \label{fig:c7c7p}}
\end{figure}

The rate of \decay{B}{X_{s}\gamma} decays is consistent with the SM and places constraints on the size of BSM contributions to the electromagnetic dipole operators $C^{(\prime)}_7$. 
The LHCb measurement of the up-down asymmetry in \decay{\Bp}{\Kp\pim\pip\gamma} decays shows that the emitted photons are polarised, but further work is needed to interpret the asymmetry. 
Moreover, measurements of $S_{\Kstar\gamma}$ from \babar and Belle are consistent with the SM expectation that $C^{}_7 \gg C^{\prime}_7$, but are not yet sufficient to rule out a sizable right-handed polarisation. 
Figure~\ref{fig:c7c7p} illustrates the current experimental constraints on $C^{\prime}_7$ after fixing all of the other Wilson coefficients to their SM values. 
A future precise measurement of $A_{\rm T}^{(2)}$ at low \qsq in the \decay{\Bz}{\Kstarz\ep\en} decay in combination with reduction on the uncertainty on $S_{\Kstar\gamma}$ would rule out a large right-handed contribution~\cite{Becirevic:2012dx}.  

\begin{figure}[!tb]
\centering
\includegraphics[width=\linewidth]{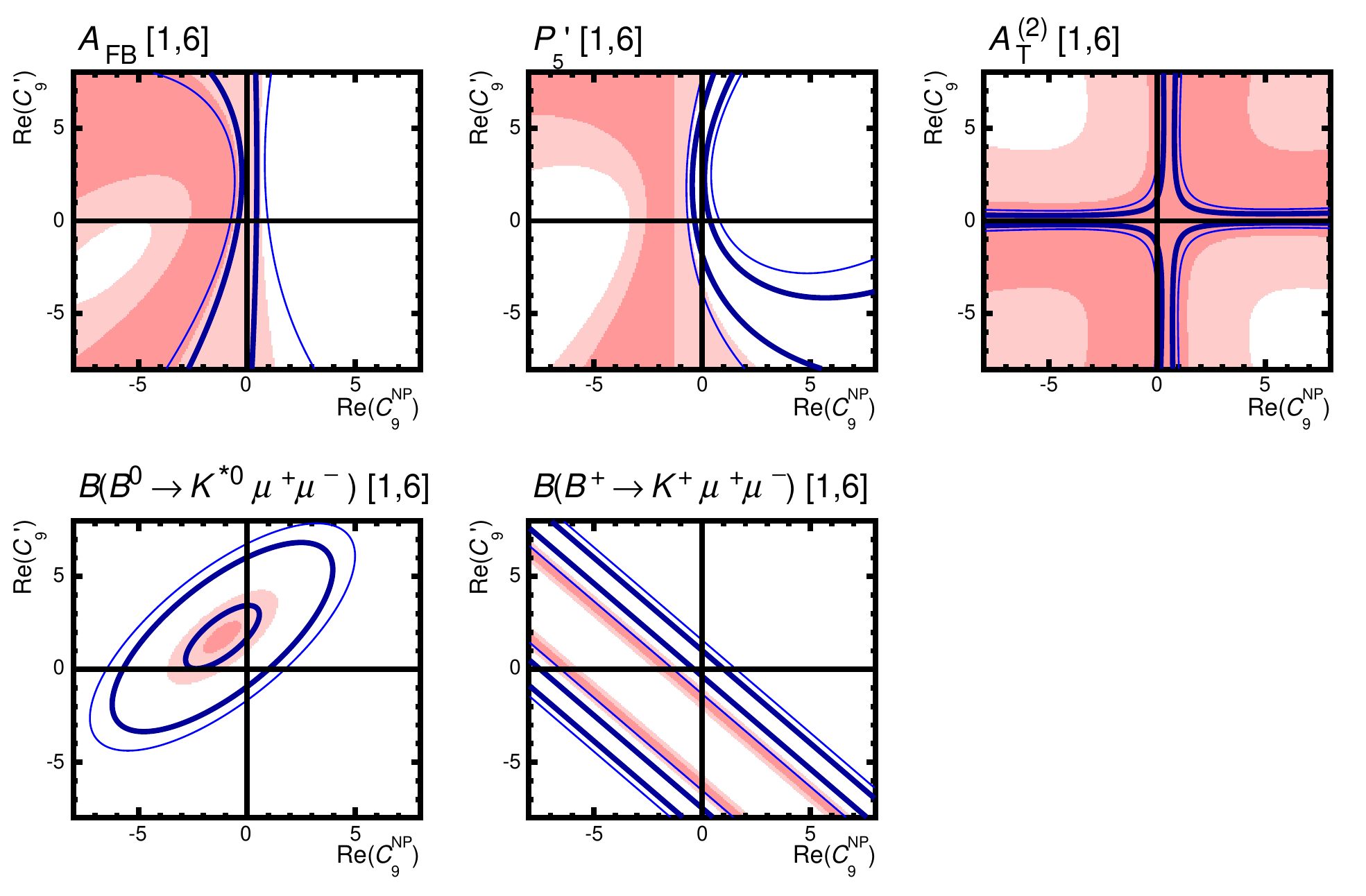}
\caption{
  Top: dependence of the angular observables $A^{}_{\rm FB}$, $P_{5}^{\prime}$ and $A_{\rm T}^{(2)}$ in the \decay{\Bz}{\Kstarz\mumu} decay in the \qsq range $1 < \qsq < 6\gev^{2}/c^{4}$ on $C_{9}^{\rm NP}$ and $C_{9}^{\prime}$;
  bottom: dependence of the branching fractions $\BF(\decay{\Bz}{\Kstarz\mumu})$ and $\BF(\decay{\Bp}{\Kp\mumu})$ in the \qsq range $1 < \qsq < 6\gev^{2}/c^{4}$;
  all computed using the EOS flavour tool~\cite{Bobeth:2010wg} with all other Wilson coefficients fixed to their SM expectations. 
  The contours indicate the 68\% and 90\%  intervals on experimental determinations of the observable. 
  The lines indicate the theoretical uncertainty on each observable. 
  \label{fig:c9c9p}}
\end{figure}

\begin{figure}[!tb]
\centering
\includegraphics[width=\linewidth]{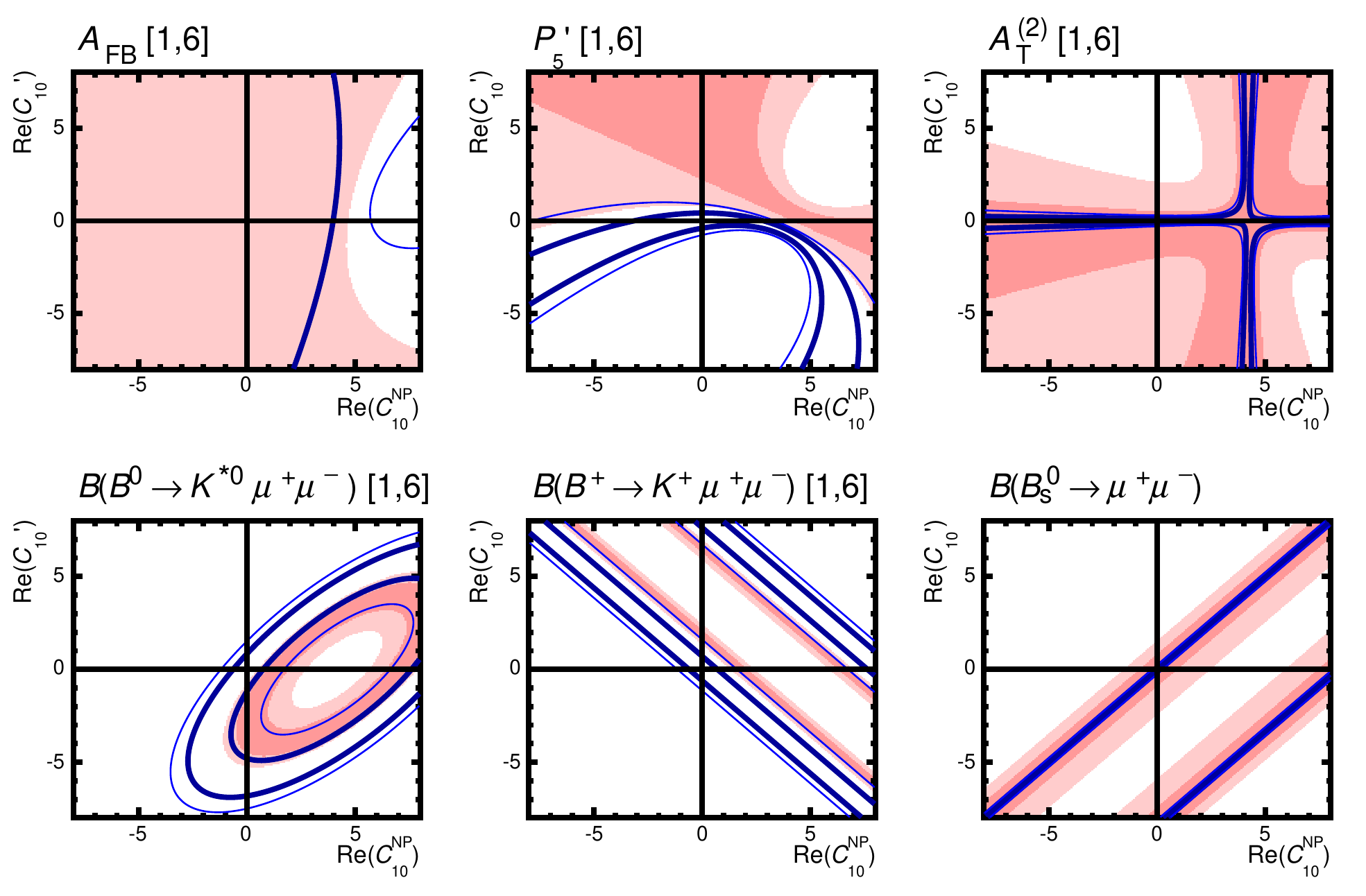}
\caption{
  Top: dependence of the angular observables $A^{}_{\rm FB}$, $P_{5}^{\prime}$ and $A_{\rm T}^{(2)}$ in the \decay{\Bz}{\Kstarz\mumu} decay in the \qsq range $1 < \qsq < 6\gev^{2}/c^{4}$  on $C_{10}^{\rm NP}$ and $C_{10}^{\prime}$;
  bottom: dependence of the branching fractions $\BF(\decay{\Bz}{\Kstarz\mumu})$ and $\BF(\decay{\Bp}{\Kp\mumu})$ in the \qsq range $1 < \qsq < 6\gev^{2}/c^{4}$ and of $\overline{\BF}(\decay{\Bs}{\mumu})$;
  all computed using the EOS flavour tool~\cite{Bobeth:2010wg} with all other Wilson coefficients fixed to their SM expectations.  
  The contours indicate the 68\% and 90\%  intervals on experimental determinations of the observables. 
  The lines indicate the theoretical uncertainty on each observable. 
  \label{fig:c10c10p}}
\end{figure}

While $C^{}_7$ and $C^{\prime}_7$ are consistent with their SM expectations, the situation with $C^{(\prime)}_{9,10}$ is more interesting. 
Measurements show a trend for the rates of the \decay{B}{K^{(*)}\mumu} and \decay{\Bs}{\phi\mumu} decays to be below their SM expectations at both low and high \qsq. 
The angular observable $P^{\prime}_5$ at low \qsq also appears to be discrepant with respect to the SM, although the  other angular observables in the \decay{\Bz}{\Kstarz\mumu} decay are reasonably consistent with their SM expectations. 
Figure~\ref{fig:c9c9p} illustrates how these measurements relate to BSM contributions to $C^{}_9$ and $C^{\prime}_9$. 
In general the data, while still consistent with the SM, are best described by a destructive BSM contribution to $C^{}_9$, which both reduces the branching fraction of the semileptonic $b \to s \mumu$ decays and also modifies the angular distribution of the \decay{\Bz}{\Kstarz\mumu} decay to be more consistent with the observed value of $P^{\prime}_5$ at low \qsq.  
Global fits to $b \to s$ data favour $C_{9}^{\rm NP} \sim -1$ with other NP parameters consistent with zero or small additional contributions to $C_9^{\prime}$ or $C^{}_{10}$~\cite{Descotes-Genon:2013wba,Beaujean:2013soa,Hurth:2013ssa,Altmannshofer:2013foa,Altmannshofer:2014rta}. 
Similar conclusions have been obtained from analyses that differ in statistical treatment, theoretical input (form factors) and in the treatment of systematic uncertainties (incorporation of power corrections). 
A reduction in the rate of $b\to s\mumu$ processes could also be achieved by a contribution to $C^{}_{10}$ with the opposite sign to the SM ($C^{\rm NP}_{10} > 0$), as shown in Fig.~\ref{fig:c10c10p}. 
However, large BSM contributions to $C^{}_{10}$ are disfavoured by the measured branching fraction of the \decay{\Bs}{\mumu} decay. 

Current fits exhibit very little sensitivity to the phases of the Wilson coefficients $C^{(\prime)}_{9}$ and $C^{(\prime)}_{10}$~\cite{Bobeth:2011gi}. 
This situation could be improved through the measurement of \CP asymmetries in the \decay{\Bz}{\Kstarz\mumu} and \decay{\Bs}{\phi\mumu} angular distributions.
Some of these asymmetries are na{\"i}ve T-odd and hence do not require sizable strong phase differences between the decay amplitudes in order to be non-vanishing. 

The measurement of $R_{K}$, Eq.~(\ref{eq:expt:rk}), raises the question of whether to consider only universal contributions to $C_{9,10}^{(\prime)}$. 
Considering flavour-dependent Wilson coefficients, the data imply
\begin{align} 
  \label{eq:Xlimit}
  & 0.7 \lesssim {\rm Re}[X^e-X^\mu] \lesssim  1.5 \, , \\
  {\rm where} \ X^\ell&=C_9^{{\rm NP} \ell} +C_9^{\prime  \ell}-(C_{10}^{{\rm NP}  \ell} +C_{10}^{\prime \ell}) \,  , \quad \ell=e, \mu \, . \nonumber
\end{align}
The anomaly could be caused by BSM physics either suppressing the dimuon mode, enhancing the (currently less constrained) dielectron channel, or both. 
Interestingly, whilst the branching fraction of the \decay{\Bp}{\Kp\mumu} decay is below the SM expectation, that of the \decay{\Bp}{\Kp\ep\en} decay is consistent with the prediction.  
The ratios $R_{K}$ and $R_{\Kstar}$ are interesting null tests of the SM because their theoretical predictions are free of hadronic uncertainties.
A measurement of $R_{\Kstar}$, or of the ratio of inclusive $b \to s \ell^+\ell^-$ branching fractions, $R_{X_s}$, in combination with the measurement of $R_{K}$ could be used to separate non-universal BSM contributions to $C^{}_{9,10}$ from those to $C^{\prime}_{9,10}$~\cite{Hiller:2014ula}.
For semileptonic decays with a ditau, or dineutrino, pair, the current data leave ample room for BSM contributions. 
In most models, however, correlations exist between the different lepton flavours and indirect bounds on the couplings can be derived from the dimuon and dielectron data.

Global analyses of the $b \to s$ data have also been explored using the ${\rm SU}(2)_{\rm L}$ invariant basis described in Sec.~\ref{sec:MIA}~\cite{Ghosh:2014awa,Hurth:2014vma,Altmannshofer:2014rta}. 
These global fits have preference for a BSM contribution to $C_{\rm LL}=C_9-C_{10}$, with left-handed muons, as opposed to $C_{\rm LR}=C_9+C_{10}$. 

Similar analyses of results on $b \to d$ decays are also possible, though the precision of the current data is not sufficient to obtain useful constraints.
Comparison of $b \to d$ with $b \to s$ data allows to search for non-MFV BSM signatures.  
Since violations of lepton universality may themselves be non-MFV signatures, if the hints for non-universality discussed above persist, it will become even more important to improve the precision of the constraints in the $b \to d$ sector.

\subsection{Limits on NP scales}

In the SM, FCNC amplitudes for $|\Delta B|=|\Delta S|=1$ transitions are suppressed because they occur through loop effects involving the weak scale ($g^2/(4\pi)^2$ and $1/M_W^2$), and also due to the smallness of the relevant CKM matrix elements, as seen in Eq.~(\ref{eq:Heff}).
As discussed in Sec.~\ref{sec:NP}, NP models may share some or all of these features.
The assumed amount of suppression of BSM amplitudes impacts on the limits on the corresponding scales.

The best fit values for $C_{9,10}^{\rm NP}$ can be interpreted in terms of a
BSM scale $\Lambda_{\rm NP}$, 
\begin{equation}
  \Lambda_{\rm NP} \times 
  \sqrt{|C^{\rm NP}_{9,10}|} \ \sim
  \begin{cases}
    ~\frac{4 \pi \,  \sqrt{2} M_W}{g e \sqrt{|V^{}_{tb} V_{ts}^*|}}   = \phantom{4}36\tev   & \mbox{(generic tree level)}, \\
    ~\frac{\sqrt{2} M_W}{ e\sqrt{ |V^{}_{tb} V_{ts}^*|}} ~ = \phantom{40}2\tev  & {\mbox{(weak loop)}}, \\
    ~ \sqrt{2} M_W/e   = 400\gev    &  \mbox{(MFV, weak loop)}.
  \end{cases}
\end{equation}
Thus, the value of $|C^{\rm NP}_9| = 1$ obtained in Sec.~\ref{sec:inter:wilson} corresponds to $\Lambda_{\rm NP}$ ranging from $\sim 400\gev$ to $\sim 36\tev$ depending on the model-dependent suppression of the BSM amplitude.
For $|\Delta B|=|\Delta D|=1$ transitions, the
CKM suppression in the SM is stronger and an analogous bound would imply a stronger constraint on $\Lambda_{\rm NP}$ in models without flavour-suppression by $\sqrt{|V_{ts}/V_{td}|} \sim 2$.

If explicit flavour factors, $\lambda\lambda^{*}$, are introduced for the BSM contribution, rare decays provide constraints on the combination $\lambda \lambda^*/\Lambda_{\rm NP}^2$.
In contrast, $\B$--$\Bbar$ mixing constrains the combination $(\lambda \lambda^*)^2/\Lambda_{\rm NP}^2$.
Similarly in the SM, the loop contribution to mixing is proportional to $G_{F} |V_{tb}^{} V_{ts}^{*}|^2$.
Due to the stronger CKM suppression and in view of the strong constraints on \CP violation in mixing, mixing bounds on $\Lambda_{\rm NP}$ are typically more powerful than those from rare decays with two important exceptions. 
The first is where the $\Delta B=1$ BSM amplitude arises at tree level but the $\Delta B=2$ BSM amplitude is loop induced, as in leptoquark models. 
Second, if the flavour suppression in the BSM model is sufficiently strong, \ie\ if $\lambda \lambda^*$ is small, rare decays provide more stringent constraints on $\Lambda_{\rm NP}$. 
It should also be emphasised that due to the different dependence on scales and couplings, combinations of measurements of $\Delta B=1$ and $\Delta B=2$ processes can be used to fix both dimensionful and dimensionless BSM parameters.

\subsection{Impact on model building}

The available data on rare \bquark~hadron decays put strong constraints on BSM contributions to the amplitudes, and this has significant impact on the range of models that can be considered.
As discussed above, BSM effects in the Wilson coefficients of the semileptonic vector ($C^{}_9$) and axial-vector operators ($C^{}_{10}$) are now limited to be no larger than roughly a third of their respective SM values, with constraints on the chirality-flipped coefficients ($C_{9,10}^\prime$) of similar size. 
The Wilson coefficients of non-SM operators (scalars, tensors) are also strongly constrained, as shown in Eqs.~(\ref{eq;Sboundmu}) and~(\ref{eq;Tboundmu}).
Note that these limits are for muons, and there is room left for possible signals in decays into other lepton species.

A common source of (pseudo-)scalar operators in BSM models is Higgs-induced penguins. 
These processes are Yukawa-dependent and hence couple differently to the different flavours of lepton.
In the minimal supersymmetric standard model (MSSM) with MFV, the introduction of a second Higgs doublet can lead to an enhancement of $C_{\rm S}^{\ell}$ and $C_{\rm P}^{\ell}$~\cite{Babu:1999hn,Bobeth:2001sq},
\begin{align}
  C_{\rm S}^\ell \simeq - C_{\rm P}^\ell \propto m_\ell \tan^3 \beta/m_A^2~.
\end{align}
Here, $\tan\beta$ is the ratio of the vacuum expectation values for the two Higgs doublets present in the theory and $m_A$ is the mass of the \CP-odd pseudoscalar Higgs. 
A sizable value of $\tan \beta$, of around fifteen, can overcome the small muon mass factor. 
The constraints on $C_{\rm S}^{\mu}$ and $C_{\rm P}^{\mu}$ arising from the
branching fraction of the \decay{\Bs}{\mumu} decay are therefore able to rule
out significant amounts of the phase-space of MSSM models~\cite{Kowalska:2013hha}.

While the overall picture is one of consistency with the SM, there are some hints of anomalies in the $b \to s$ data, which if taken at face value exhibit quite interesting features:
a) a preference for sizable NP in leptonic vector couplings, with axial-vector NP not larger in magnitude;
b) a preference for lepton non-universality;
c) the possibility of right-handed currents.
If future measurements substantiate any of these, there will be huge implications for model building. 
To accommodate such features requires models that go beyond the most common and simple solutions to the hierarchy problem. 
For instance, b) and c) directly imply a non-MFV flavour sector, while b) also suggests lepton-flavour violation~\cite{Glashow:2014iga}.
It should be noted that the significant level of non-universality hinted at in the data, if confirmed, would rule out many proposed SM extensions including the MSSM with R-parity conservation~\cite{Martin:1997ns}.

Feature a) has inspired model-building with $Z^\prime$-extensions to the SM. 
As argued in Sec.~\ref{sec:NP}, many models with possibilities for large $Z$-penguins predict the hierarchy between NP in axial-vector and vector couplings to be the other way around. 
However, one popular approach that prefers large vector NP is the ${\rm SU}(3) \times {\rm SU}(3) \times {\rm U}(1)$, or ``3-3-1'', model~\cite{Gauld:2013qja,Buras:2013dea}, which can also accommodate feature c). 
A model which can explain all of a), b) and c) is the gauged $L_\tau$--$L_\mu$ extension of the SM~\cite{Fox:2011qd,Altmannshofer:2014cfa}. 
A variant of the latter with an additional Higgs-doublet~\cite{Crivellin:2015mga} can also explain the $2.5\sigma$ hint from CMS of Higgs boson decay to $\tau^\mp \mu^\pm$~\cite{CMS:Htaumu}. 
In this model there are constraints on the ratio of the mass of the $Z^\prime$ and of its coupling, $m_{Z^\prime}/g^\prime$, in the few \tev range.

Leptoquark models offer a natural framework to accommodate lepton non-universality. 
Choosing an ${\rm SU}(2)_{\rm L}$-triplet leptoquark that couples to muon doublets one obtains a model that induces at tree level~\cite{Gripaios:2014tna,Hiller:2014yaa}
\begin{equation}
  \label{eq:LQ:model}
  C_9^{\rm NP\mu}=-C_{10}^{\rm NP \mu}=\frac{\pi}{\alpha_{e}} \frac{\lambda_{s\mu}^*\lambda^{}_{b\mu}}{V_{tb}^{} V_{ts}^*}
  \frac{\sqrt{2}}{2M^2G_F}~,
\end{equation}
where $M$ is the mass of the leptoquark.
With benchmark values to explain the measurement of $R_{K}$ given in Eq.~(\ref{eq:expt:rk}), $M^2 \simeq \lambda_{s\mu}^*\lambda^{}_{b\mu} \, (48 \tev)^2$.
Viable flavour structures for the leptoquark couplings $\lambda_{q \ell}$ can arise in models with partial compositeness~\cite{Gripaios:2014tna}.
Dineutrino operators are induced, which enhance the branching ratios of \decay{B}{K^{(*)} \nu \bar{\nu}} and \decay{B}{X_s \nu \bar{\nu}} decays by a few percent. 
Corrections of a few percent to $C^{}_7$ can also arise.
The relation between axial-vector and vector coupling of Eq.~(\ref{eq:LQ:model}) can be broken if more than one leptoquark is introduced.

The leptoquarks may be searched for through their decays
\begin{equation}
  \phi^{ 2/3} \to t\ \nu ~, \quad
  \phi^{-1/3} \to b\ \nu ~, t\ \mu^- ~, \quad
  \phi^{-4/3} \to b\ \mu^- ~,
\end{equation}
where the final state particles must be from different generations since leptoquarks carry two generational indices.
Such distinctive signatures will however only be visible in experiments in the leptoquarks are sufficiently light.

\section{Summary and outlook}
\label{sec:summary}

Run~I of the LHC has led to a substantial improvement in precision in several key observables among rare decays of \bquark~hadrons.
Particularly notable are the first observation of the very rare \decay{\Bs}{\mumu} decay and the wide range of kinematic observables now studied in \decay{\Bz}{\Kstarz\mumu} decays.
The results remain broadly consistent with the SM, but deviations are present at an intriguing level of significance.
In the light of this situation, and the high sensitivity to BSM physics provided by rare \bquark~hadron decays, it is essential to continue to improve the precision.

This presents challenges for both theory and experiment.
One important task for theory is to reduce uncertainties that arise due to hadronic effects in the decays.
This includes improving precision in decay constants and form factors, for example through refined lattice QCD calculations.  
In the case of \decay{b}{s\ell^+\ell^-} decays, other important issues are to understand power corrections and resonance contributions.
To address these issues will require improved computations as well as detailed studies of data on specific observables and fits.
The data will allow not only to determine certain hadronic parameters, but will also provide consistency checks and allow theoretical methods to be refined.

On the experimental side, the situation varies between \bquark~hadron decay modes.
As shown in Table~\ref{tab:b2ll}, the precision of branching fraction measurements of \decay{\Bds}{\ell^+\ell^-} decays is not yet at the level of the theory predictions; only for \decay{\Bs}{\mumu} is it within a factor of 3.
Improvements in these measurements can be achieved with a combination of experimental facilities.
Data taking at the LHC, in Run~II and beyond, will provide large increases in the yields of dimuon decays.
The experiments will benefit both from the increased cross-section for production of \bquark~quarks, which is expected to scale approximately linearly with the centre-of-mass energy, and from the high luminosity.
The increased yields will allow not only improved branching fraction measurements, and a determination of their ratio, but also a study of the effective lifetime in \decay{\Bs}{\mumu} decays~\cite{DeBruyn:2012wk}.
The LHC experiments may also be able to improve on the existing limits for the dielectron and ditau modes, but substantial improvement is more likely to be possible at the Belle~II experiment~\cite{Aushev:2010bq} and at a very high luminosity $e^+e^- \to Z$ factory~\cite{Zhao:2002zk,Gomez-Ceballos:2013zzn}, which have an experimental environment more suitable to these modes.

Radiative \bquark~hadron decays provide the potential for significant future improvement in the knowledge of right-handed contributions to the $b \to s\gamma$ dipole amplitude.
To achieve this it will be necessary to use all of the methods most sensitive to the photon polarisation, since they provide complementary information~\cite{Becirevic:2012dx}.
These methods include time-dependent asymmetries in \decay{\Bz}{\KS\piz\gamma} and \decay{\Bs}{\phi\gamma} decays, up-down asymmetries in \decay{\Bp}{\Kp\pip\pim\gamma} decays, and angular asymmetries in \decay{\Bz}{\Kstarz\epem} decays.
Improved searches for \CP violation in both inclusive and exclusive processes, as well as tests of MFV from $\BR(\decay{\Bz}{\rhoz\gamma})/\BR(\decay{\Bz}{\Kstarz\gamma})$ will also be important.
There are excellent prospects for progress in all of these areas at both LHCb, including its upgrade~\cite{LHCb-PAPER-2012-031}, and Belle~II.

Data from the LHC have allowed dramatic progress in experimental investigations into semileptonic \decay{\bquark}{\squark \ell^+\ell^-} decays.
The precision of the measurements of the differential branching fractions as a function of \qsq is now good enough that the focus is primarily on asymmetries, including \CP asymmetries, isospin asymmetries, and lepton universality violating differences in rates. 
With the increasing yields available in decay modes such as \decay{\Bz}{\Kstarz\mumu}, full angular analyses are expected to become mandatory.
It will also be possible to fit for \CP violating angular observables.
The reconstruction of semileptonic modes with dielectron and ditau pairs is more challenging.
The immediate objectives are to search for violations of lepton universality in a wide range of modes, including \decay{\Bz}{\Kstarz\epem} and \decay{\Bs}{\phi\epem}, and to make first efforts at searching for the ditau modes.
In addition, high luminosity \epem experiments are expected to reach interesting sensitivity to the \decay{\B}{K^{(*)}\nu\bar{\nu}} decays, which provide complementary information on BSM physics since they are sensitive to other Wilson coefficients.

The main goal of ongoing investigations into rare \bquark~hadron decays is to first uncover evidence of BSM physics, and then to deduce its nature.
As discussed in this review, the quantity and quality of the data being produced by the LHC provide exciting prospects.
The hints of BSM physics in the data analysed so far have led to new directions in model-building with the desired phenomenological features.
To exploit fully the data that will be available from the LHC and from future experiments will require increased effort to identify and interpret patterns in the short-distance coefficients.
This will provide fantastic opportunities to learn about the problems and puzzles in fundamental physics that remain in the SM.
This challenge can best be confronted by both theory and experiment in collaboration.

\subsection*{Note added in proof}

After the completion of this review, preliminary results from the full analysis of the LHCb Run~I data on the angular distributions of the $B^0 \to K^{*0} \mu^+ \mu^-$ decay became available~\cite{LHCb-CONF-2015-002}. 
Careful study will be required before the detailed implications of these results are understood; however, consistency with the earlier results based on a subset of the data~\cite{LHCb-PAPER-2013-019,LHCb-PAPER-2013-037} is seen and therefore the interpretation of the new data is likely to proceed along similar lines as discussed in this review.

\section*{Acknowledgements}

TB and TG thank the other members of the LHCb collaboration for the enjoyable and productive teamwork that has led to many of the results discussed in this review.
This work is supported in part by the Royal Society (TB), the European Research Council under FP7 and the Science and Technology Facilities Council (TG), and by the Bundesministerium f\"ur Bildung und Forschung (BMBF) and the DFG Research Unit FOR 1873 ``Quark Flavour Physics and Effective Field Theories'' (GH).
The authors thank Christoph Bobeth and Michal Kreps for useful comments.

\bibliographystyle{LHCb}
\bibliography{references,LHCb-PAPER}

\end{document}